\documentclass[aps,twocolumn,preprintnumbers,superscriptaddress,showpacs,nofootinbib,amsmath, amssymb]{revtex4-1}

\usepackage{graphicx}
\usepackage{subfigure}
\usepackage{hyperref}
\usepackage{color}
\usepackage[table]{xcolor}
\usepackage{boldline,multirow}
\usepackage{booktabs}
\usepackage{pifont}
\usepackage{mathrsfs}
\usepackage{soul}
\usepackage[utf8]{inputenc}
\usepackage{multirow}

\usepackage{color}
\definecolor{rosso}{cmyk}{0,1,1,0.4}
\definecolor{rossos}{cmyk}{0,1,1,0.55}
\definecolor{rossoc}{cmyk}{0,1,1,0.2}
\definecolor{blu}{cmyk}{1,1,0,0.3}
\definecolor{blus}{cmyk}{1,1,0,0.6}
\definecolor{bluc}{cmyk}{1,1,0,0.1}
\definecolor{verde}{cmyk}{0.92,0,0.59,0.25}
\definecolor{verdec}{cmyk}{0.92,0,0.59,0.15}
\definecolor{verdes}{cmyk}{0.92,0,0.59,0.4}
\definecolor{bviolet}{rgb}{0.54, 0.17, 0.89}
\hypersetup{colorlinks,bookmarksopen,bookmarksnumbered,citecolor=bviolet,
linkcolor=bviolet,pdfstartview=FitH,urlcolor=bviolet}

\def\apj{Astrophys.\ J.}

\def\apjl{Astrophys.\ J.\ Lett.}
\def\aap{Astron.\ \& Astrophys.}

\def\araa{Ann.\ Rev.\ Astron.\ \& Astrophys.}

\begin{document}

\preprint{UCI-TR-2020-03}

\title{Entering the Era of Dark Matter Astronomy? \\ Near to Long-Term Forecasts in X-Ray and Gamma-Ray Bands}
\author{Dawei Zhong}\email{dzhong4@uci.edu}
\affiliation{Department of Astronomy, Xiamen University, Xiamen, Fujian 361005, People’s Republic of China}
\affiliation{Center for Cosmology, Department of Physics \& Astronomy, University of California, Irvine, Irvine, CA 92697, USA}
\author{Mauro Valli}\email{mvalli@uci.edu}
\affiliation{Center for Cosmology, Department of Physics \& Astronomy, University of California, Irvine, Irvine, CA 92697, USA}
\author{Kevork N.\ Abazajian}\email{kevork@uci.edu}
\affiliation{Center for Cosmology, Department of Physics \& Astronomy, University of California, Irvine, Irvine, CA 92697, USA}

\begin{abstract}
We assess Galactic Dark Matter (DM) sensitivities to photons from annihilation and decay using the spatial and kinematic information determined by state-of-the-art simulations in the Latte suite of Feedback In Realistic Environments (FIRE-2). For kinematic information, we study the energy shift pattern of DM narrow emission lines predicted in FIRE-2 and discuss its potential as DM-signal diagnosis, showing for the first time the power of symmetric observations around $l=0^{\circ}$. We find that the exposures needed to resolve the line separation of DM to gas by XRISM at $5\sigma$ to be large, $\gtrsim 4$~Ms, while exposures are smaller for Athena ($\lesssim 50$~ks) and Lynx ($\lesssim 100$~ks). We find that large field-of-view exposures remain the most sensitive methods for detection of DM annihilation or decay by the luminosity of signals in the field of view dominating velocity information. The $\sim$4 sr view of the Galactic Center region by the Wide Field Monitor (WFM) aboard the eXTP mission will be highly sensitive to DM signals, with a prospect of $\sim 10^5$ to $10^6$ events from the 3.5~keV line in a 100 ks exposure, with the range dependent on photon acceptance in WFM's field of view. We also investigate detailed all-sky luminosity maps for both DM annihilation and decay signals -- evaluating the signal-to-noise for a DM detection with realistic X-ray and gamma-ray backgrounds -- as a guideline for what could be a forthcoming era of DM astronomy. 
\end{abstract}

\maketitle

\section{\label{sec:level1}Introduction}
Today we know that Dark Matter (DM) accounts for $\sim 85$\% of the matter amount present in our observable Universe, and constitutes $\sim 1/4$ of the total inferred cosmological energy budget~\cite{Aghanim:2018eyx}. While its presence has been playing a pivotal role in our understanding of the formation and evolution of structures in the Universe, the origin of DM remains essentially unknown~\cite{Bertone:2018xtm}. It is known that new degrees of freedom beyond the ones pertaining to the Standard Model of Particle Physics must exist for a viable DM candidate~\cite{feng2010,Bertone:2010zza}. In large classes of DM models, the DM candidates' coupling with Standard Model particles in the microscopic theory, as well as the early-time mechanism for DM production in the early Universe, provide a late-time mechanism for photon signatures due to the DM particle's decay or annihilation, visible in the X-ray \cite{Abazajian:2001vt} to gamma-ray \cite{Baltz:2008wd}. 

Two recent signatures of photons from DM have generated significant interest: an excess of gamma-ray photons toward the Milky Way's Galactic Center as detected by the Fermi-LAT satellite, consistent with expected DM spatial profiles, with intensity peaking in the 2-3 GeV range \cite{goodenough09,hooper11}, and the detection of an unidentified line around 3.5~keV in numerous observations (see review in Ref.~\cite{Abazajian:2017tcc}). The line was originally discovered in the Perseus Cluster, stacked clusters and Andromeda (M31) \cite{bulbul14,boyarsky14}. These compelling signatures beg the question of what will emerge from increasingly deeper and more robust observations of the high-energy photon sky, whether these previous results are confirmed as due to DM or not. In addition to the primary tree-level mechanisms for photon production from DM particle annihilation scenarios, monochromatic photons are produced in many models at the loop-level \cite{Abazajian:2011tk}. These lines can therefore also be used for the so-called ``DM spectroscopy'' that we study here, though they do not constitute the dominant emission. Eventually, internal states in DM can also produce narrow emission features of interest \cite{Finkbeiner:2009mi,Finkbeiner:2014sja}. 

In order to spot the annihilation or decay of DM particles in galactic halos, high-energy photons stand out as the most promising indirect messengers~\cite{Bergstrom:1997fj}, leaving the original information on the spectrum and morphology of DM signals essentially unaltered~\cite{Bringmann:2012ez,Gaggero:2018zbd}. The Milky Way (MW) halo is known to significantly dominate over extra-galactic signals in both X-ray \cite{Abazajian:2006jc} and gamma-ray observations \cite{Abazajian:2010sq}. Therefore, it represents a primary target for our study. Our knowledge of the DM halo in the MW and in the whole Local Group plays a very important role in assessing the sensitivity for a possible detection of high-energy photons from DM. In the last decade, several investigations with N-body simulations have been performed, clearly identifying the Galactic Center (GC) as the most interesting region where to look for DM annihilation/decay in the gamma-ray band~\cite{diemand07,springel08,Springel:2008cc,kuhlen08,Pieri:2007ir,Pieri:2009je,kamionkowski10}. The possibility of boosted galactic DM signals in virtue of the presence of substructures has also been received increasing attention in the community~\cite{Gao:2011rf,Sanchez-Conde:2013yxa,Moline:2016pbm,Ando:2019xlm,Hutten:2019tew}. 

In the photon data collected by the Fermi-LAT satellite, the spacecraft collaboration has confirmed the appearance of a roughly spherical pattern at few GeV within tens of degrees in longitude and latitude around the GC~\cite{TheFermi-LAT:2015kwa,TheFermi-LAT:2017vmf}. This was interpreted from early analyses as potentially the first signature of GeV-scale DM annihilating in the MW halo~\cite{goodenough09,hooper11,abazajian12,Daylan:2014rsa,Calore:2014xka}. Such a compelling picture is challenged by other reasonable astrophysical scenarios, e.g. unresolved sources~\cite{Abazajian:2010zy,abazajian12,Abazajian:2014fta} or cosmic-ray (CR) physics~\cite{Gaggero:2015nsa,Carlson:2016iis}, and possibly questioned by the absence of similar signals in the subhalos hosting MW dwarf satellites, see e.g.~\cite{Ahnen:2016qkx,Keeley:2017fbz,Hoof:2018hyn} (but see also \cite{Bonnivard:2014kza,Ullio:2016kvy,Ichikawa:2017rph,2020arXiv200211956A} for caveats).
The DM interpretation remains of interest, see e.g. the recent discussion in~\cite{Leane:2019uhc,Chang:2019ars,Leane:2020nmi,Buschmann:2020adf} and the analysis in Ref.~\cite{Karwin:2019jpy}. However, strong constraints on DM may now also be derived~\cite{Abazajian2020} when adopting Galactic bulge, nuclear stellar cluster and gas emission templates that substantially improve the description of the GC region~\cite{Macias:2016nev,Bartels:2017vsx,Macias:2019omb}.

At the same time, spacecraft missions like the {\it Chandra X-ray Space Telescope}, {\it XMM-Newton} and {\it NuSTAR} have also provided us a quite interesting angle on fundamental physics via X-ray astronomy. Of particular significance, the aforementioned detection of a $\sim\!3.5$~keV emission line from the stacked spectral analysis of nearby galaxy clusters~\cite{bulbul14}, as well as Andromeda and Perseus clusters~\cite{boyarsky14}. The line has also been detected or indicated, at different levels of significance, toward the  GC by {\it XMM-Newton} ~\cite{Boyarsky:2014ska}, in the Perseus Cluster by the {\it Suzaku Telescope} \cite{Urban:2014yda}, in deep sky observations by {\it NuSTAR} \cite{Neronov:2016wdd} and by {\it Chandra} \cite{Cappelluti:2017ywp}, toward the GC out to $\sim$10$^\circ$ \cite{Boyarsky:2018ktr}, and in the MW Galactic Bulge limiting window \cite{Hofmann:2019ihc}. While the interpretation of these findings in terms of astrophysical emission may be still considered under debate~\cite{bulbul14,Jeltema:2014qfa,Gu:2015gqm,Shah:2016efh}, as of today, the series of indications of the 3.5~keV line in the X-ray sky represent one of the most intriguing claims for uncovering the particle nature of DM~\cite{Abazajian:2017tcc,Boyarsky:2018tvu}. 
In this regard, the recent theoretical study in Ref.~\cite{speckhard16} set up the stages for a possible era of DM spectroscopy. There, the authors entertained the possibility of distinguishing emission lines on the basis of the spectral information carried out by DM, showing how it would be expected to come very different from the contributions tracing baryonic matter. Further investigation has been put forward with N-body simulations in Ref.~\cite{Powell:2016zbo} in order to assess more accurately the feasibility of the proposal, and the possible gauging of the DM 3.5~keV signal via the Micro-X rocket spectrometer~\cite{Micro_X}.

In this work, we are motivated by the exciting prospects for DM astronomy of our own MW's DM halo, namely by advanced X-ray missions such as Micro-X \cite{Adams:2019nbz}, XRISM~\cite{XRISM}, eXTP~\cite{Zhang:2016ach}, Athena~\cite{athena} and Lynx \cite{LynxTeam:2018usc}, as well as future gamma-ray facilities like e-ASTROGAM~\cite{eastrogam}, AMEGO~\cite{McEnery:2019tcm}, HERD~\cite{Zhang:2014qga,Huang:2015fca}, GAMMA-400~\cite{Egorov:2017nyt}, and CTA~\cite{Bigongiari:2016amk, Acharya:2017ttl}. As such, we reassess the emissivity of Galactic DM based on state-of-the-art hydrodynamic simulations of MW-like galaxies, exploiting the Latte suite of Feedback In Realistic Environments (FIRE-2)~\cite{wetzel16, kimmel17, sanderson2018, hopkins18}. We present an up-to-date investigation on the science case represented by DM spectroscopy using the outcome of FIRE-2 simulations, and providing full-resolution results as well as forecasts relevant for forthcoming X-ray missions.  

We also reassess energy resolution, field of view (FOV) and exposure in the sensitivity to DM signals from our own MW's DM halo for the nearest-term missions, XRISM, which will have sensitivity to velocity information, and eXTP, which will have great sensitivity to the luminosity signal due to the large FOV of the Wide Field Monitor (WFM)~\cite{Hernanz:2018llr} aboard the observatory. 
Two other developments have allowed for a new assessment: first, increasingly precise and accurate galaxy formation simulations of high resolution that include star formation and hydrodynamics, and, second, enhanced understanding of foregrounds to DM signals allow for a more detailed forecast of the expected signal and its robustness over astrophysical emission. Observations of the high-energy sky will take a new leap forward with microcalorimetry in X-ray astronomy \cite{XRISM,athena,LynxTeam:2018usc,Adams:2019nbz} as well as enhanced energy resolution in the hard X-ray to gamma-ray \cite{eastrogam,McEnery:2019tcm,Zhang:2014qga,GAMMA400}. We show in this paper that the current status of the theoretical forecast of DM emission surpasses the short to medium-term future sensitivity of high-energy photon observatories. This allows for an exploration of DM luminosity and spectroscopy on the sky by idealized instruments beyond the limitations of current or near-term experimental observatories. And, maybe most importantly, the forecast signal and signal-to-background we explore here can help guide future mission design and development. These topics are what we study in this paper, with special attention to the spectroscopic DM Doppler shift (velocity spectroscopy), DM dispersion line broadening, as well as the expected DM emissivity. Most importantly, we show that large FOV observations of narrow-energy features remain the most effective method of detection of a DM line as well as its differentiation from an astrophysical source. Such large FOV observations will occur with the WFM instrument aboard eXTP~\cite{Zhang:2018edu}.

This paper is organized as follows: In Section~\ref{sec:level2} we present the most relevant findings in our work concerning the possibility of carrying out DM spectroscopy using the velocity field of DM particles stored in the Latte suite of FIRE-2 simulation. We discuss in detail the essential steps for a clear DM-signal diagnosis under the assumption of a net emission line detection. Section~\ref{sec:level3} is devoted to reassess carefully all-sky luminosity maps for both DM annihilation and decay signals in X-ray and gamma-ray bands exploiting once again the outcome of FIRE-2. Most importantly, we evaluate the expected signal-to-noise ratio for DM against realistic X-ray and gamma-ray background modeling. Finally, we summarize our results in Section~\ref{sec:level4} and report further details on the analysis in appendix~\ref{app:level1}~-~\ref{app:level2}.

\begin{figure*}
    \begin{center}
    \includegraphics[width=0.45\textwidth]{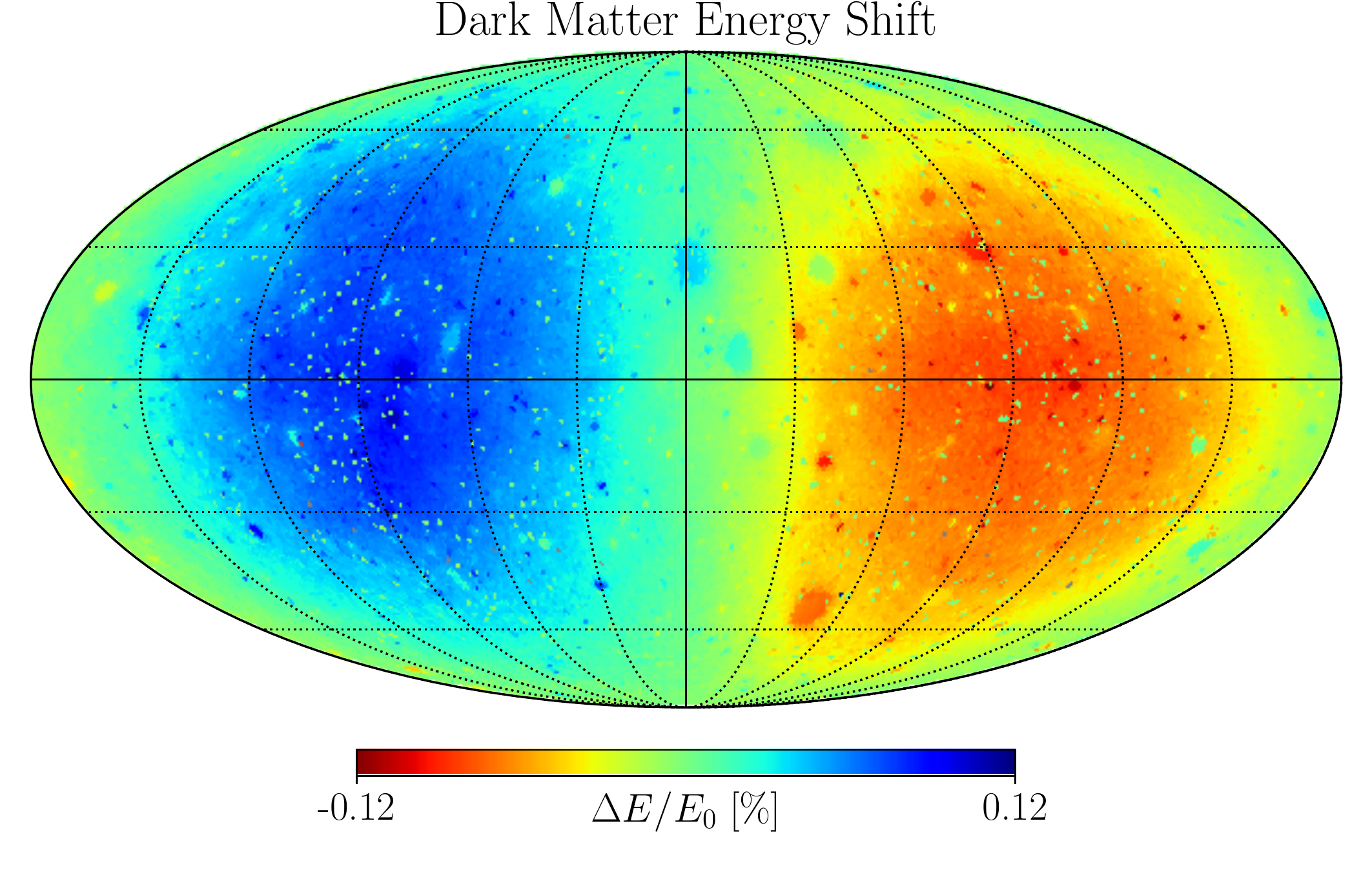}
    \hspace{10pt} 
    \includegraphics[width=0.45\textwidth]{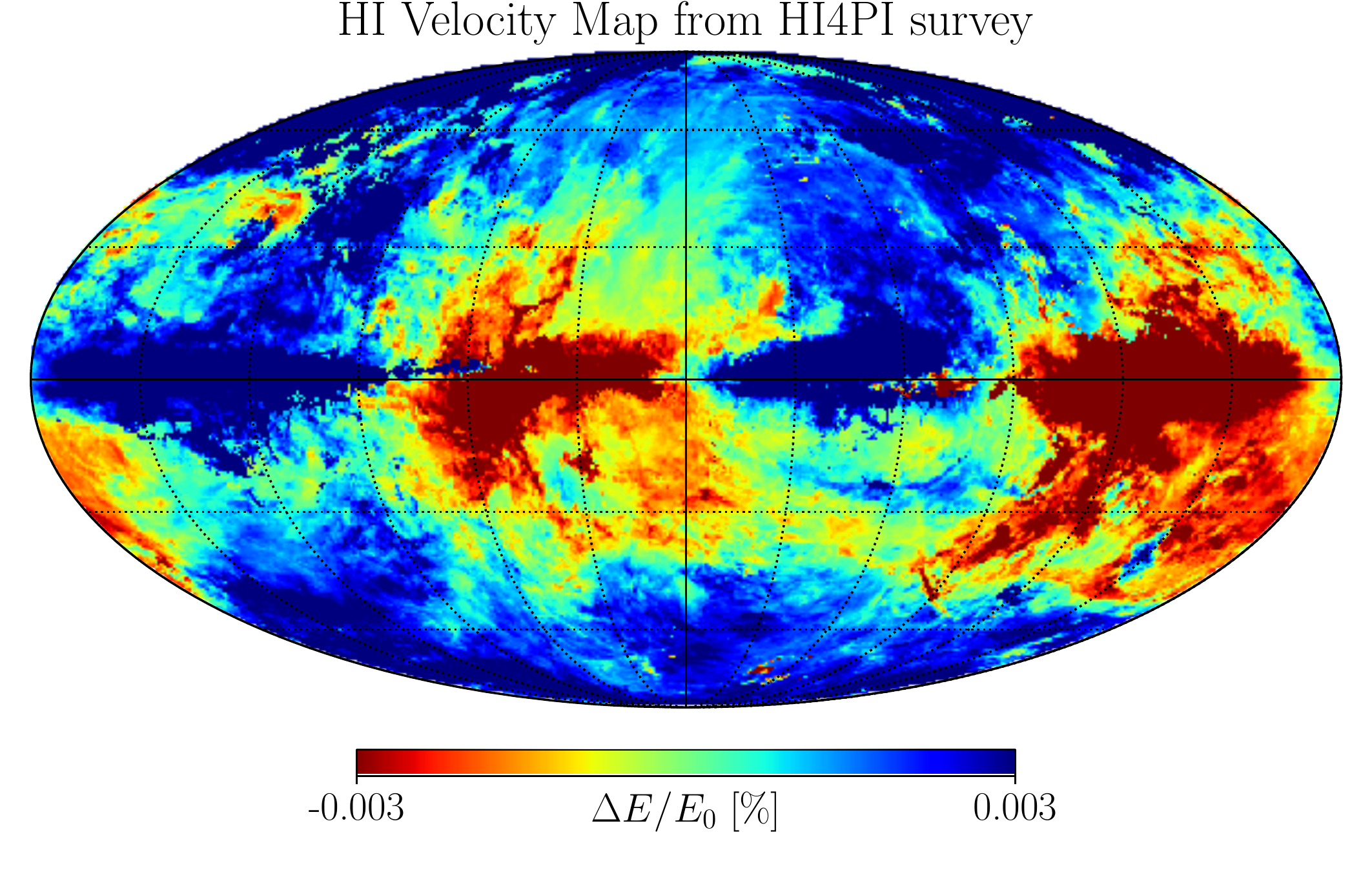}
    \end{center}
    \caption{Left panel: The energy shift pattern of a DM narrow emission line estimated from the DM l.o.s. velocity dispersion extracted from the FIRE-2 \textit{m12i} zoom-in simulation. The red/blue pattern in the skymap corresponds to l.o.s. where the DM signal gets redshifted/blueshifted as a result of the relative motion with respect to the observer. Right panel: The energy shift pattern of Galactic 21cm HI emission line from HI4PI survey. Similarly, the red/blue part denotes the redshift/blueshift region for the HI emission line. As elaborated in the text, the distinctive energy shift pattern exhibited by the two maps (left and right panel) could be used for an accurate diagnosis of the origin of an unresolved line. }
        \label{fig:spec}
\end{figure*}

\section{\label{sec:level2} Doppler shift of Dark Matter}
Let us start considering the following question: if an emission line in the X-ray or in the gamma-ray wavelength would be detected, how would it be unambiguously related to DM physics?
Lines associated to high-energy gamma-rays -- say photons of tens to hundreds of GeV~\cite{Weniger:2012tx}, or even more energetic -- would not have an obvious astrophysical counterpart: together with the morphological information of the signal, they would indeed provide a smoking gun for the discovery of particle DM~\cite{Bringmann:2012ez}. However, in the case of a line detection involving softer photons, i.e. in the keV~\cite{Abazajian:2001vt} and in the MeV~\cite{Bartels:2017dpb} energy range -- where multiple astrophysical emission lines appear -- discarding the background-only hypothesis can be a much harder task to accomplish.

As originally discussed in Ref.~\cite{speckhard16}, a promising route to uncover an emission line in favor of DM may consequently rely on the analysis of its Doppler shift, i.e. via a study of DM velocity spectroscopy.
For the purpose, let us recall that the differential flux expected from DM annihilation/decay up to overall constants reads:\footnote{See appendix~\ref{app:level1} for some more detail.}
\begin{equation}
   \label{eq:flux_DM}
    \frac{d \Phi_{\chi}}{d \Omega d E} \propto \frac{\mathcal{L}_{\chi}}{4\pi} \frac{dN_{\chi}}{dE}  \ .
\end{equation}
In the above, $\mathcal{L}_{\chi}$ stands for the ``DM (relative) luminosity'' (often quoted in literature also as $D-$ and $J-$factor~\cite{Bergstrom:1997fj} for DM decay and annihilation, respectively) while $dN_{\chi}/dE$ is the photon energy spectrum from DM final states. Note that $\mathcal{L}_{\chi}$ is sensitive to the macroscopic properties of DM, while the photon spectrum is instead related to the details of DM particle interactions: the two are in principle factorized from each other (but see, e.g., Refs.~\cite{Boddy:2017vpe,Petac:2018gue} for exceptions due to velocity-dependent DM annihilation). However, if we would have at our disposal a detector with an exquisite energy resolution (say, at least, at the level of $\mathcal{O}(0.1\%)$~\cite{speckhard16}), then the above factorization would cease to be valid; e.g., for the case of decaying DM one should rewrite the RHS of Eq.~\eqref{eq:flux_DM} as~\cite{speckhard16,Powell:2016zbo}:
\begin{equation}
   \label{eq:flux_DM_shift}
   \frac{1}{4 \pi} \int_{\rm los} d\vec{s}  \ \rho_{\chi}[r(\vec{s}\,)] \int d \mathcal{E} \frac{dN_{\chi}}{d \mathcal{E}} \mathcal{K}\left(\mathcal{E}, E, \sigma_{\rm los}[r(\vec{s}\,)] \right) \ ,
\end{equation}
where $ \rho_{\chi}$ is the DM density profile, while $\mathcal{K}$ is a spectral kernel that is function of energy and also of the line-of-sight (l.o.s.) velocity dispersion of DM, $\sigma_{\rm los}(r)$. Assuming a line with optimal detector resolution, Eq.~\eqref{eq:flux_DM_shift} improves the estimate in Eq.~\eqref{eq:flux_DM} encoding now the energy shift of the original spectrum $dN_{\chi}/dE$ due to the relative motion of DM with respect to us. For example, for an emission line generated by DM decay with rate $\Gamma_{\chi}$ and mass $m_{\chi}$, one would expect the original spectrum $dN_{\chi}/dE = \delta(E - m_{\chi}/2) $; then, approximating $\mathcal{K}$ with a Gaussian profile, the expected differential flux of photons from DM, $d \Phi_{\chi}/(d \Omega d E)$, would be explicitly given by: 
\begin{equation}
    \label{eq:DM_decay_Gauss}
    \frac{\Gamma_{\chi}}{(2\pi)^{3/2}m_{\chi}} \int_{\rm los} d\vec{s}  \ (\rho_{\chi}/ \sigma_{\rm los}) \, e^{-\frac{1}{2}\left(\frac{m_{\chi}-2 E}{ \sigma_{\rm los} m_{\chi}} \right)^2} \ .
\end{equation}
As we will discuss in this section, the spectral energy shift due to DM relative motion could have a crucial impact in the future prospects for the interpretation of the 3.5~keV emission line, whose detection today is established at high statistical significance.

In order to study DM spectroscopy, as a starting point of our investigation we analyzed the outcome of cosmological zoom-in baryonic simulations of MW-mass galaxies from the Latte suite of FIRE-2\footnote{For the interested reader, visit \href{https://fire.northwestern.edu/}{https://fire.northwestern.edu/.}}~\cite{hopkins18}. These simulations represent the state-of-the-art in the field, involving the so-called ``GIZMO gravity plus hydrodynamics'' code in meshless finite-mass mode, allowing for an accurate inclusion of hydrodynamic effects in N-body studies, see~\cite{hopkins15}. In particular, the underlying physics implemented in FIRE-2 simulations contains the modeling of star formation, feedback, and also cooling/heating processes in a multi-phased interstellar medium (ISM)~\cite{hopkins18}. 
The whole cosmological simulation initially contained several individual DM halos within a box of 85.5 Mpc in length, evolved under the assumption of standard $\Lambda\text{CDM}$ cosmology. 
A suite of MW-mass halos present in the cosmological simulation have been rerun in order to get zoom-in regions at redshift $z=0$~
\cite{wetzel16, kimmel17, sanderson2018}. 

From the aforementioned suite, the \textit{m12i} \cite{wetzel16}, \textit{m12f} \cite{kimmel17} and \textit{m12m} \cite{hopkins18} realizations have been already used in synthetic surveys\footnote{See, e.g., Gaia DR2-like samples collected \href{https://girder.hub.yt/\#collection/5b0427b2e9914800018237da/folder/5b211e42323d120001c7a813}{here}.}
as they constitute the \textit{Latte} suite halos, which reproduce to a good approximation properties of the Galaxy, e.g. $M_{\text{halo}} \sim 10^{12}M_{\odot}$~\cite{hopkins18}. In the present study, we chose to focus on the \textit{m12i} halo that also well reproduces the stellar and gas mass inferred for the MW~\cite{sanderson2018}, and provides a concrete playground for the analysis of the distribution of DM in the Galaxy. Note that the \textit{m12i} simulation contains more than $7 \times 10^7$ high-resolution DM particles with mass of $ M_{p} = 3.5 \times 10^4 M_{\odot}$ within the zoom-in region (600 kpc), centered around the MW-mass halo. Each DM particle in the simulation is characterized by both 3-dimensional spatial position $\vec{x}=(x,y,z)$ and velocity $\vec{v}=(v_{x}, v_{y}, v_{z})$ defined in the GC reference frame. In our analysis we restrict to DM particles within a sphere of about about the virial radius of the \textit{m12i} galaxy (i.e. roughly 300~kpc), resulting in a total of about $3 \times 10^6$ particles.

We used HEALPix~\cite{HEALPix,healpy} with a resolution index set to 6 as the optimal choice to produce a detailed skymap for DM, ending up with a fair number of particles in each pixel for a meaningful statistics.
For the computation of the line-of-sight velocity of DM, $v_{\text{los}}$, relative to the observer, we  evaluated the corresponding energy shift for each DM particle, $\Delta E/E_{0} = -v_{\text{los}}/(c+v_{\text{los}})$, being $E_{0} = m_{\chi}/2$ the characteristic energy of the emission line. For a given l.o.s., we  binned the result in order to infer the spectral function $\mathcal{K}$ from the simulation. We adopted a Voigt profile and performed a fit to each of the histograms obtained in correspondence to the line-of-sight pixel in the DM skymap. We used the centroid of the fitted Voigt profile to estimate the shift in the photon spectrum $\Delta E/E_{0}$.\footnote{For a few cases, a single Voigt profile was not an optimal choice to fit the histogram obtained for $\Delta E/E_{0}$. The estimate of the centroid of the line for these cases is discussed in Appendix~\ref{app:level2}.} 

In the left panel of Fig.~\ref{fig:spec} we present the all-sky inferred energy shift $\Delta E/E_{0}$, reported in percentage with a Mollweide projection.
The dipole structure characterizing the DM $v_{\rm los}$ skymap obtained is the result of the DM halo being static while the observer (at the position of the Sun) co-rotates with the Galactic disk. Therefore, in the rest frame of the observer, the bulk of DM particles in the window $l=0-180^{\circ}$ are moving towards ($v_{\text{los}}<0$) while those in the sky region $l=180^{\circ}-360^{\circ}$ are moving away ($v_{\text{los}}>0$), resulting in a corresponding dipole blueshift or redshift of the DM emission energy, respectively. In the same panel, peaks in the $v_{\rm los}$ of DM are visible in correspondence to the clustered inhomogeneities of the map due to the presence of subhalos.

In sharp contrast to what extracted from the FIRE-2 simulation for DM, we show in the right panel of Fig.~\ref{fig:spec} the energy-shift map due to the $v_{\rm los}$ of the Galactic HI gas as characterized in the analysis of Ref.~\cite{HI4PI}.\footnote{The velocity distribution of gas and stars in the FIRE-2 simulation does not match well the observed HI map, highlighting a current limitation of state-of-the-art N-body implementations of baryonic physics. However, we checked explicitly that the DM result shown in Fig.~\ref{fig:spec} is not altered by this fact, being consistent also with the outcome of DM-only simulations.} In order to obtain the velocity gas map shown, we adopted the 21~cm HI emission data of HI4PI~survey\footnote{The interested reader can find the map at this \href{http://cdsarc.u-strasbg.fr/viz-bin/qcat?J/A+A/594/A116}{link}.} and analyzed the spectra obtained from the dataset in the ancillary material provided by the collaboration. In particular, each of these spectra corresponds to a profile of HI brightness temperature versus the HI line-of-sight velocity and we fitted the HI emission line using for simplicity a single Gaussian function. Despite our simplistic modeling, the best-fit result allowed us to obtain a reasonable estimate of the centroid of the 21~cm line emission. The resulting HI energy-shift pattern reported in the right skymap of Fig.~\ref{fig:spec} is morphologically complex, and very different from the DM one. Indeed, the ``blue-red-blue-red'' structures in the all-sky map are connected to the motion of HI gas, correlated with the rotation of the Galactic disk. In addition to the morphological information emerging from Fig.~\ref{fig:spec}, we wish to note that the $v_{\rm los}$ of the HI gas happens to be much smaller than the one inferred for DM: hence, the energy shift of an emission line associated to the motion of the HI gas would be typically smaller than the one expected from DM particles. 

\begin{figure*}
    \centering
    \includegraphics[width=0.45\textwidth]{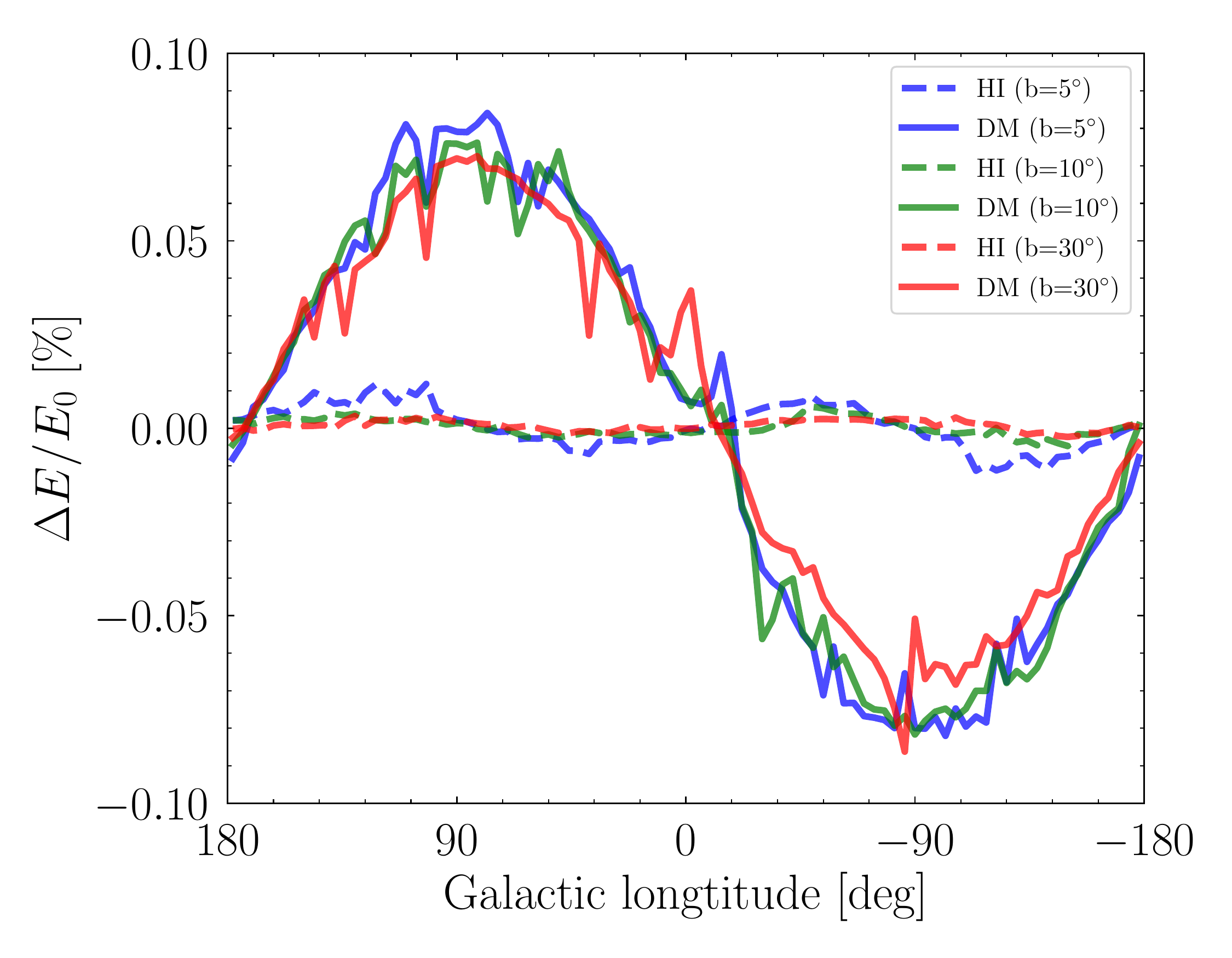}
    \hspace{10pt} \includegraphics[width=0.45\textwidth]{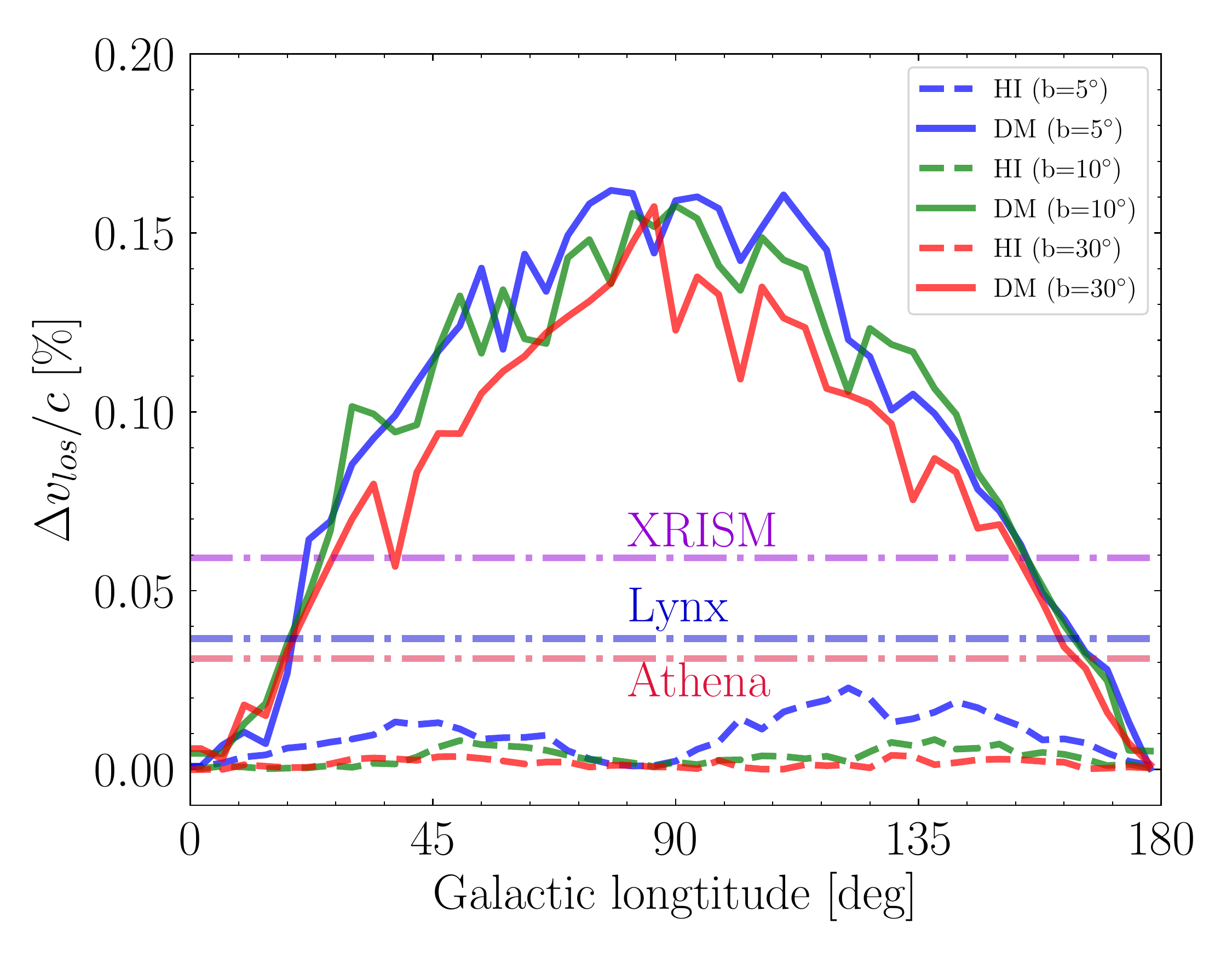}
    \caption{Left panel: The relative line shift $\Delta E/E_{0}$ versus Galactic longitude $l$ at Galactic latitude $b = 5^{\circ},10^{\circ}$ and $30^{\circ}$. Solid lines denote DM shift while dashed lines are for HI gas shift. Direct comparison between these two cases shows that at $\l \sim 90^\circ$ the Doppler shift from DM can be clearly distinguished from the astrophysical one. Right panel: the separation of two emission lines as given in Eq.\eqref{eq:delta_DeltaE} observed around $l=0^{\circ}$ at fixed $b = 5^{\circ},10^{\circ}$ and $30^{\circ}$. The horizontal dash-dotted lines show the expected energy resolution ($1\sigma$, {\it not FWHM}) of future X-ray telescopes at 3.5~keV. Though the separation, $\Delta v_{los}/c$, is within a factor of $\sim$2 of the nominal energy resolution, we estimate that the centroids of the DM versus gas lines may be separated at $>\! 5 \sigma$ with XRISM exposures of 3.9 to 15 Ms, Athena exposures of 10 to 52 ks, and Lynx exposures of 23 to 110 ks, whether a very aggressive or a much more conservative estimate of the expected background is considered, see Table~\ref{tab:xraymission}.}
    \label{fig:vcurve}
\end{figure*}

The differential motion of gas and DM is quite striking in the left panel of Fig.~\ref{fig:vcurve}, where we display the variation of the energy shift $\Delta E/E_{0}$ versus Galactic longitude at fixed Galactic latitude $b = 5^{\circ},10^{\circ}\text{ and }30^{\circ}$. Clearly, the magnitude of a line-emission energy shift imprinted by the relative motion of DM stands out over the one coming from the kinematics of the HI gas. Besides, the DM longitude profile shown is approximately symmetric around $l=0^{\circ}$, and yields similar shifts $\Delta E/E_{0}$ for the different Galactic latitudes considered. This is again in sharp contrast with the case of the HI gas, which follows a more complex pattern. Eventually, we wish to highlight that the DM longitude profile of Fig.~\ref{fig:vcurve} presents spiky irregularities in virtue of the contribution of substructures along the line-of-sight: these can further blueshift/redshift spectra as already seen in the Doppler shift map of Fig.~\ref{fig:spec}. 

Let us now focus more on the possibility of performing a diagnosis of the origin of an emission line, as for the case of the 3.5~keV detection, using DM spectroscopy. For the purpose, let us consider the possibility of a simultaneous detection of a DM line in coincidence with an astrophysical background emission happening at the same characteristic energy $E_{0}$: assuming a unity ratio of the two components, how would we really discern one from the other? 
With the aim of providing a quantitative answer to this question, in the right panel of Fig.~\ref{fig:vcurve} we report the difference of the line-of-sight velocity measured in the sky at same latitude ($b = 5^{\circ}, 10^{\circ}, 30^{\circ}$ again for illustrative purposes), but $\pm l$ pair of longitude coordinates:
\begin{equation} 
    \label{eq:delta_DeltaE}
    \Delta v_{los}/c \equiv \Delta E / E_{0}\big|_{\bar{l}} - \Delta E / E_{0}\big|_{-\bar{l}} \ , 
\end{equation}
with $\bar{l} \in$~[$0^{\circ},180^{\circ}$].

Clearly, an interesting opportunity emerges from the inspection of this quantity: Fig.~\ref{fig:vcurve} shows how two observations of the sky at same latitude but performed at a pair of Galactic longitudes close to $\bar{l} = 90^{\circ}$ would provide a very promising diagnosis of the line origin under scrutiny. Note that here we assumed the HI gas kinematics to yield the dominant source of background expected for the analysis. While this assumption may turn out to be rough (or even inaccurate given the partial information we have about emitting gas in the Galaxy, see~\cite{26Al, fbout1, fbout2}), it would be hard to imagine that any other astrophysical source of Galactic origin for the line could be degenerate with the spectroscopic identification of DM due to the expected co-rotation within the MW disc~\cite{speckhard16}. A possible notable exception to this argument may be encountered in the gaseous halo inferred at the outskirts of the MW~\cite{2011ApJ...737...22A}, whose degree of rotation remains uncertain, see~Ref.~\cite{2017ApJ...849..105L}. However, independently from the present characterization of the MW diffuse X-ray halo, the spectroscopy of a DM emission of photons of energy $E_{0} \gtrsim$ 2~keV  would wash out the potential degeneracy with this astrophysical background.

Hence, in the right panel of Fig.~\ref{fig:vcurve} we show the DM longitude profile related to the velocity difference introduced in Eq.~\eqref{eq:delta_DeltaE}. Interestingly, the prediction reported -- agnostic a priori on the characteristic spectral energy $E_{0}$ -- turns out to be above the prospects of detection of future X-ray campaigns as the one of Athena, XRISM and Lynx for the 3.5~keV emission line, see also Table~\ref{tab:xraymission}. From Fig.~\ref{fig:vcurve} we can therefore conclude that, in the future, performing two observations at two symmetric points around $l=0^{\circ}$ with same Galactic latitude, say $b<30^{\circ}$, would allow us to gain a unique clue on the interpretation of the 3.5~keV line. 

The velocity spectroscopy method can be employed with a number of proposed telescopes. The X‐ray microcalorimeter on the upcoming XRISM has a planned energy resolution 5~eV (FWHM) at 6~keV \cite{XRISM}. The Athena X-ray Integral Field Unit (X-IFU) would provide a high resolution measurement at the energy range of $0.2-12.0\text{ keV}$ \cite{Barret:2018qft}. Based on the \textit{TESSIM} simulation, the energy resolution of X-IFU will be approximately 2.5 eV (FWHM) near 7 keV \cite{Barret:2016ett}. The X-ray microcalorimeter on the Lynx X-ray Observatory is proposed to also have very high energy resolution, with the Main Array and the Enhanced Main Array having uniform energy resolution $3\text{ eV}$ (FWHM) in $0.2-7.0\text{ keV}$ \cite{LynxTeam:2018usc}. The missions' energy resolutions are relatively independent of energy in the range of interest for our approximation of required exposures. However, the effective area is more highly dependent of energy, and we use that for 3.5 keV in the respective missions' current science cases \cite{XRISM,Barret:2018qft,LynxTeam:2018usc}, as listed in Table~\ref{tab:xraymission}. To estimate the signal, we used canonical Navarro-Frenk-White (NFW)~\cite{Navarro:1995iw} parameters as $\rho_{H} = 0.4\text{ GeV cm}^{-3}$ (or $\sim 1.05 \times 10^{7} M_{\odot} \text{ kpc}^{3}$) and $R_{H} = 16\text{ kpc}$~\cite{MW_NFW} (see also discussion in section~\ref{sec:level3}).  

Let us therefore consider the detectability of the line separation of the 3.5~keV emission line in the MW. We adopt $\Delta v_{los}/c  = 0.15$ and consider the difference of two line centroid at $l=\pm 90^{\circ}$ to be $\Delta E\approx 5.25\text{ eV}$ (c.f. Fig. \ref{fig:vcurve}). We list the energy resolution and the separation between two observations at $l=\pm 90^{\circ}$ in Table~\ref{tab:xraymission}, and we show the relative energy resolution $\Delta E/E_{0}$ in Fig.~\ref{fig:vcurve} (right panel). By integrating counts, the center of the line's energy can be increasingly more precisely determined. We estimate the uncertainty in  $\delta E = C(R)\sigma_\mathrm{eff}/N_s$, where $\sigma_\mathrm{eff}$ is the effective energy resolution of the instrument, $N_s$ is the number of signal photons, and $C(R)=\sqrt{1+4R}$ is a factor determined by the signal-to-background, $R\equiv N_\mathrm{bkg}/N_s$. Assuming a range from an optimistic vanishing-background model ($R\ll 1$), to a conservative high-background flux model $(R\approx 1)$,  we find that the velocity separation of the lines can be accomplished by XRISM at $5\sigma$ one-sided separation with an exposure between 3.9 to 15 Ms. Athena and Lynx have much greater sensitivity to the line separation due to their higher energy resolution and effective areas, with required $5\sigma$ exposures of 10 to 52 ks for Athena, and 23 to 110 ks for Lynx, depending on the estimate of the background. Our estimates are consistent with the separation sensitivity that was considered in Speckhard et al. \cite{speckhard16} ($\approx\! 3.6\sigma$ at $l= 20^\circ$), but here we consider the standard high statistical threshold requirement of 5$\sigma$ and greater sensitivity availed by the $l=\pm 90^\circ$ symmetric observations about $l=0$.

\begin{table}
\centering
\begin{ruledtabular}
\begin{tabular}{cccc}
\textit{X-ray Mission} & {\bf Lynx }  & {\bf Athena } & {\bf XRISM}\\ 
\hline
\hline\\
Energy Resolution [eV] (FWHM) & 3 & 2.5 & 5\\[0.1cm]
Effective Area [cm$^2$] (@ 3.5 keV) & 4000 & 6000 & 250 \\ \\
\hline \\
\multicolumn{3}{r}{ \textit{\underline{Exposures for \boldmath$5\sigma$ detection} \ \ \ \ \ \ \ \,}}  & \\ \\
{Low X-ray background case}    &  23 ks  &  10 ks  &  3.9 Ms \\[0.1cm]
{High X-ray background case}   &  110 ks  &  52 ks   &  15 Ms \\[0.3cm]
\end{tabular}
\end{ruledtabular}
\caption{For proposed instruments' energy resolutions, fields of view and effective areas of these respective missions, we provide the minimum exposure time for $5\sigma$ line separation due to velocity differentiation between foreground emission and the 3.5~keV line case in pair observation at $l=\pm 90^\circ$ and $|b|=20^\circ$ for a low X-ray background ($N_\textrm{bkg}\ll N_s$) and high X-ray background cases ($N_\textrm{bkg} \approx N_s$). 
}
\label{tab:xraymission}
\end{table}

\begin{figure}
    \begin{center}
    \includegraphics[width=0.45\textwidth]{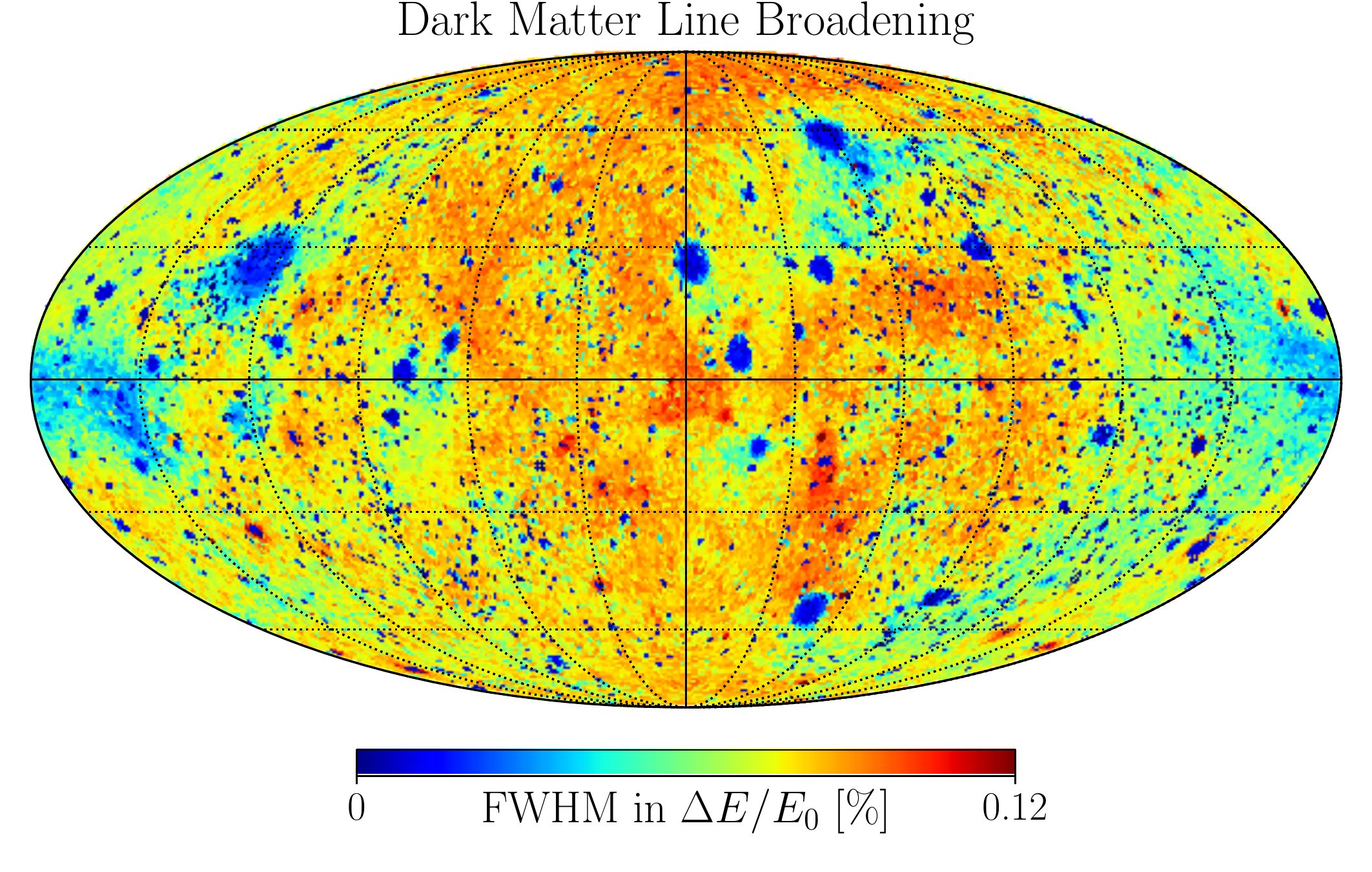}
    \end{center}
    \caption{The full-width-half-maximum (FWHM) of DM narrow emission line estimated from the DM velocity dispersion in each line-of-sight. The colder, blue portions have narrow observed lines from the dominance of cold subhalo contributions, while the warmer, red portions are due to the higher velocity dispersion in the parent halo. }
    \label{fig:width}
\end{figure}

While upcoming X-ray campaigns could have a possible measurement of DM Doppler shift effects with long exposure times,  the next generation of gamma-ray telescopes do not have enough spectral resolution for this purpose:  the required energy resolution emerging from Fig.~\ref{fig:vcurve} should be at least $\mathcal{O}(0.1\%)$ and would need instruments beyond those currently proposed. For example, the spectral resolution of e-ASTROGAM would be about $\Delta E/E_{0} >0.4\%$ at $0.1-10\text{ MeV}$ \cite{DeAngelis:2016slk}; the energy resolution of HERD would correspond instead to $1\%$ for gamma-rays beyond 100 GeV~\cite{Zhang:2014qga,Huang:2015fca} and similar resolution would be at hand for GAMMA-400 \cite{Egorov:2017nyt}, and CTA~\cite{Bigongiari:2016amk, Acharya:2017ttl}; eventually, AMEGO would have an energy resolution slightly below $1\%$ for energies $\lesssim 2 \text{ MeV}$ (and quite worse, $\sim 10\%$, at the $\text{ GeV})$ \cite{McEnery:2019tcm}. 

Looking forward the eventual possibilities with spectroscopic information,  lines from X-ray or gamma-ray photons would not only be Doppler shifted, but have an intrinsic velocity dispersion. That dispersion is modeled by the simulation, and we show the intrinsic width of the line(s) on the sky in Fig.~\ref{fig:width}. We describe the method for producing this map in Appendix~\ref{app:level2}. The required energy resolution is at the $\ll \! 0.1\%$ level. Note the dark blue (cold) regions in the velocity dispersion are due to colder substructure bound to the parent halo. These structures could eventually be probed by very high energy-resolution observatories, providing a look into the cold, early forming structure in our own Galaxy's DM halo. 

\begin{figure*}
    \begin{center}
    \includegraphics[width=0.45\textwidth]{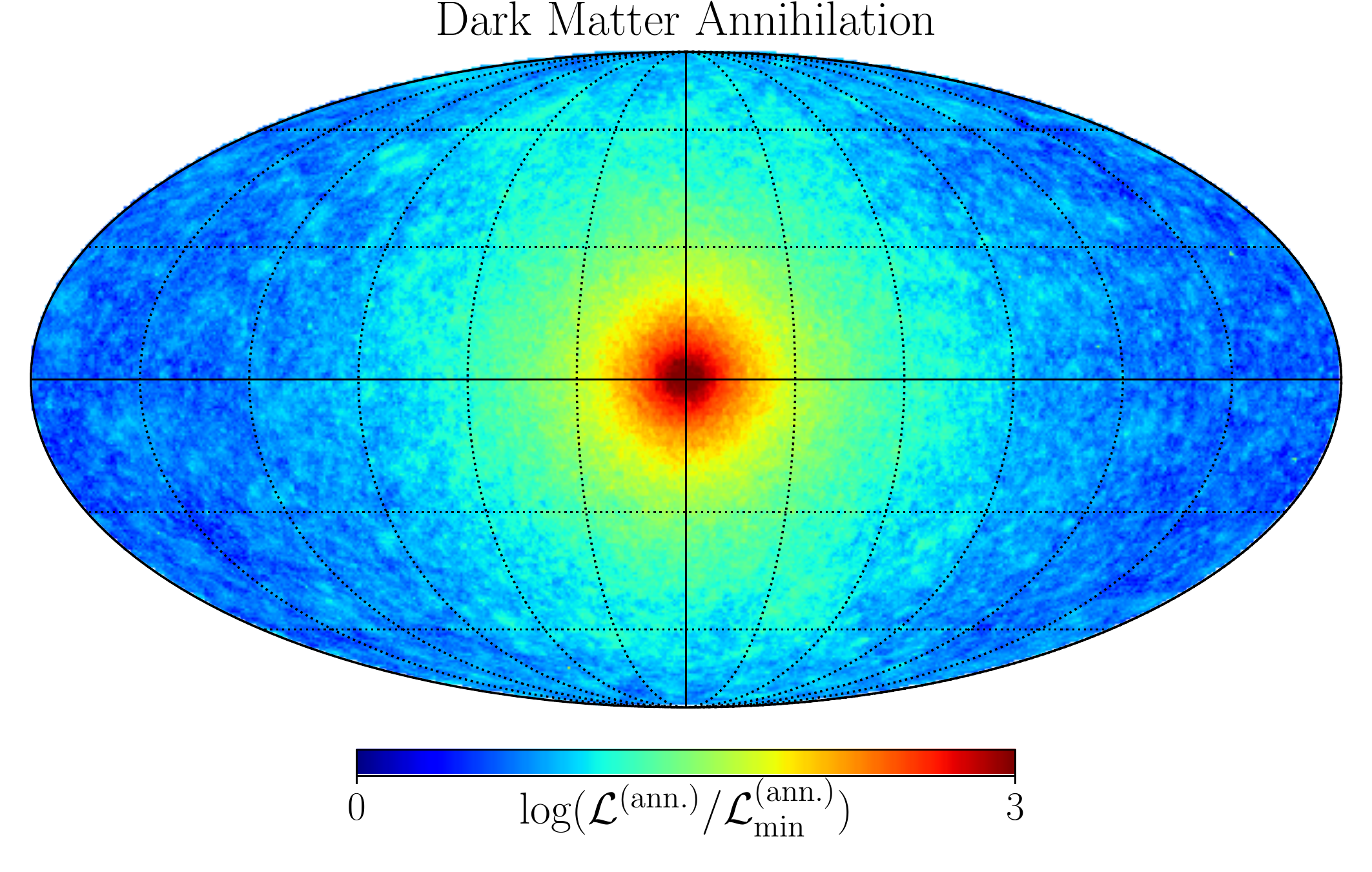}
    \hspace{10pt} \includegraphics[width=0.45\textwidth]{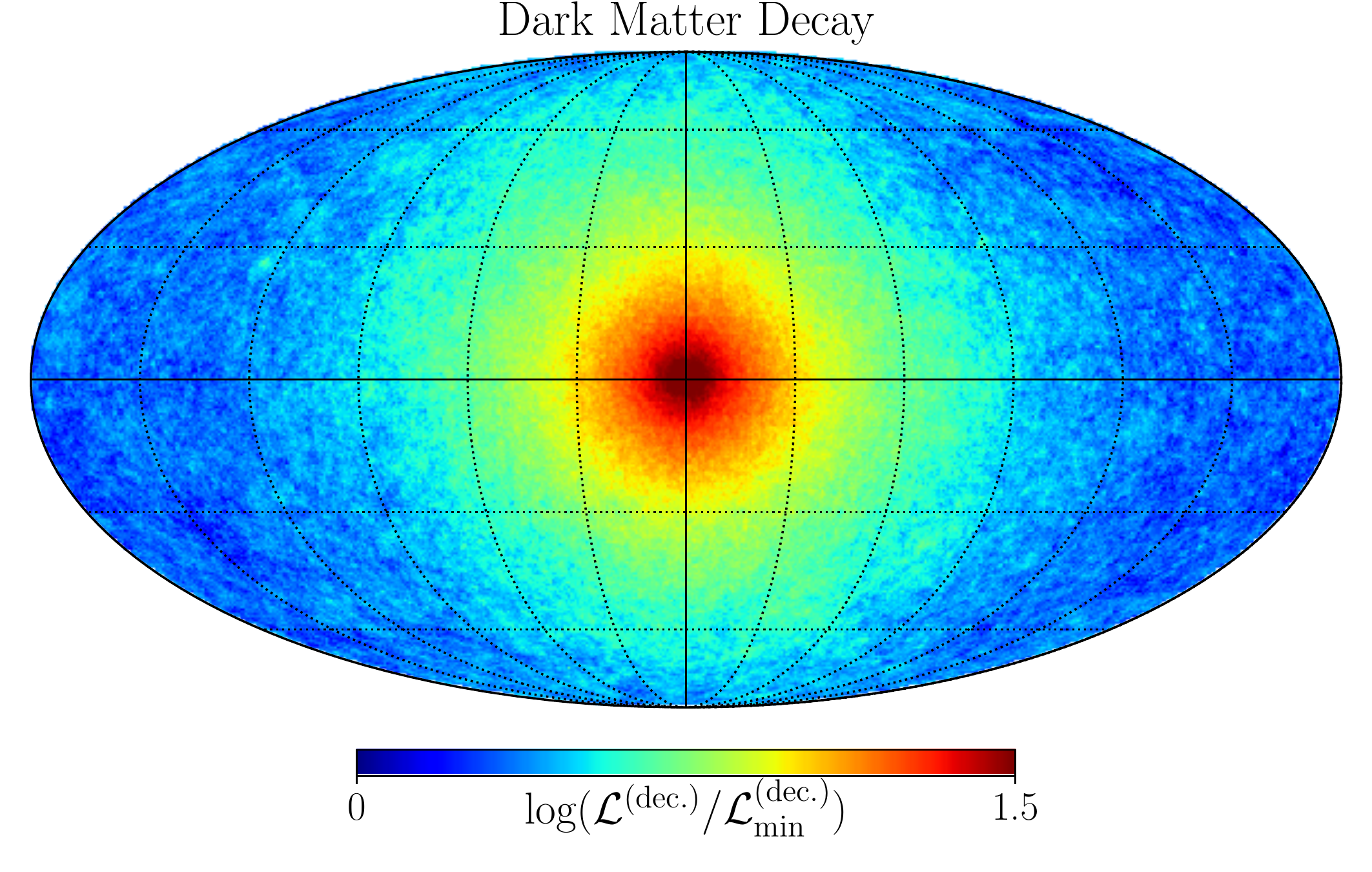}
    \end{center}
    \caption{Left panel: The Galactic DM annihilation luminosity distribution from an observer at the Sun position (at $R_{\odot} = 8.2$~kpc) in the FIRE-2 \textit{m12i} simulation.  Right: The analogous Galactic luminosity map for the case of DM decay. See text for more details on the full-sky prediction of DM luminosity from FIRE-2.  }
    \label{fig:lumin_map}
\end{figure*}

\section{\label{sec:level3}Dark matter luminosity in the X-ray and gamma-ray Sky}

\begin{figure}[!htb!]
    
    \centering
    \includegraphics[width=0.45\textwidth]{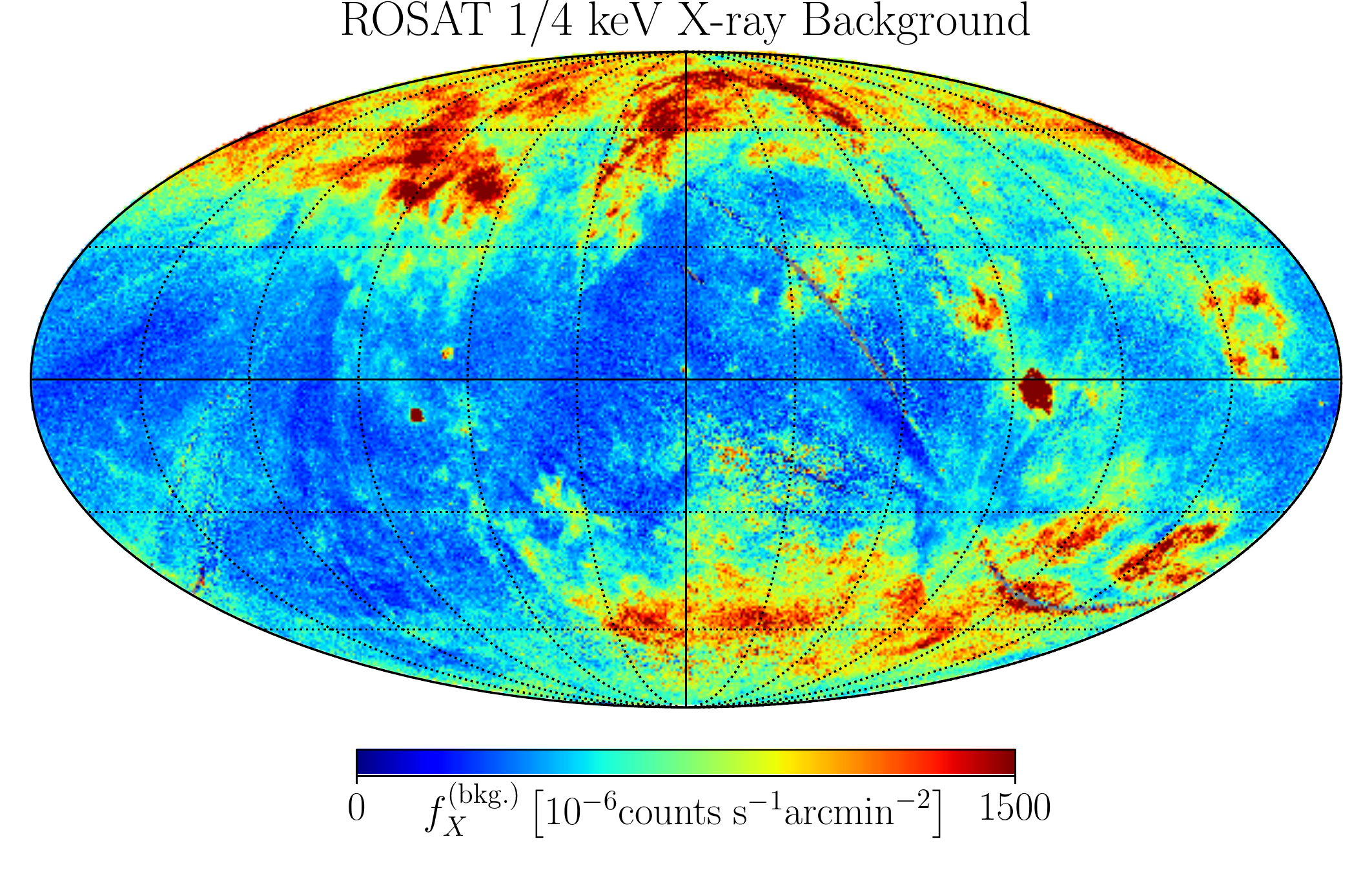}
    \vspace{3pt}
    \includegraphics[width=0.45\textwidth]{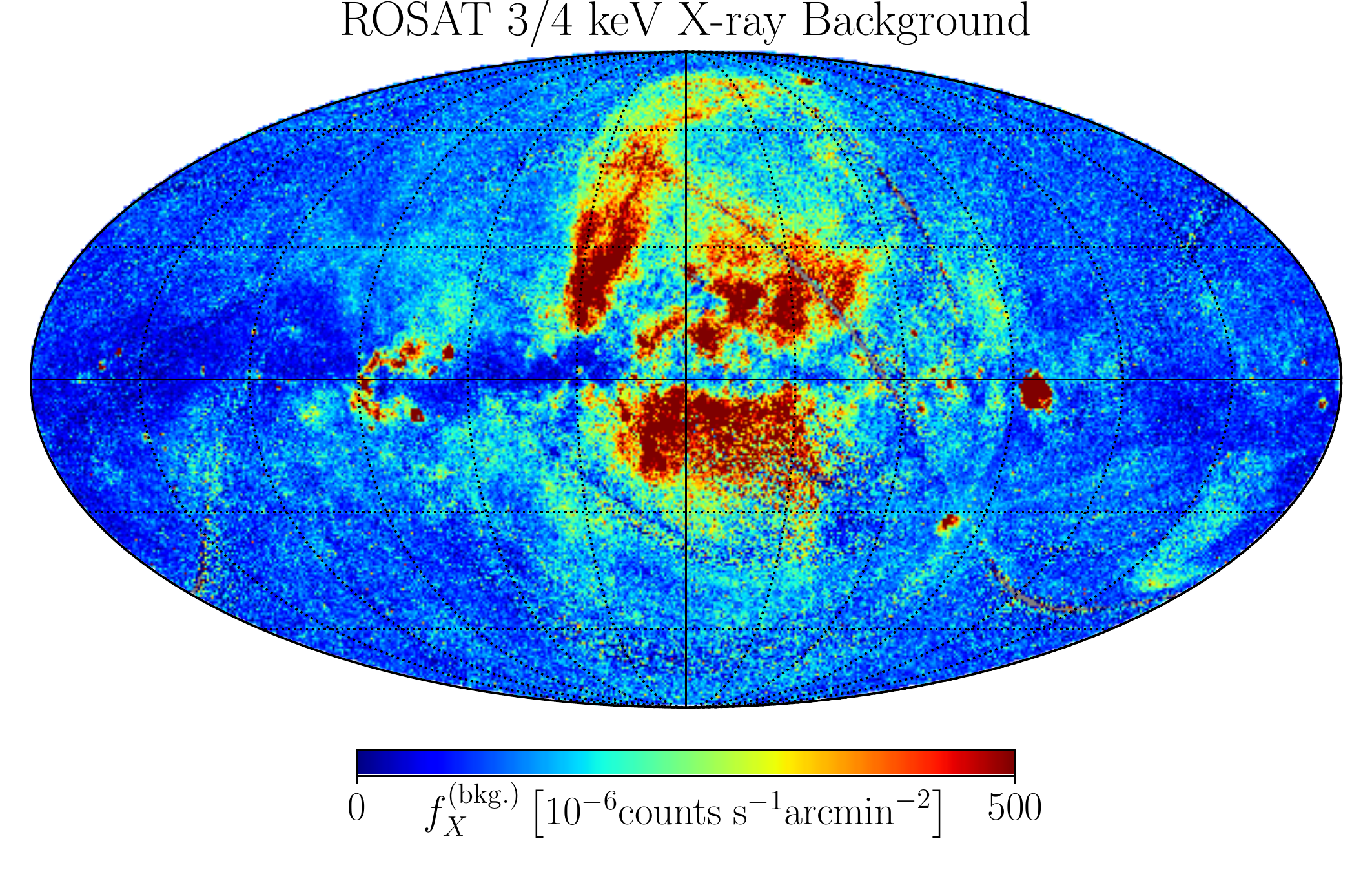}
    \vspace{3pt}
    \includegraphics[width=0.45\textwidth]{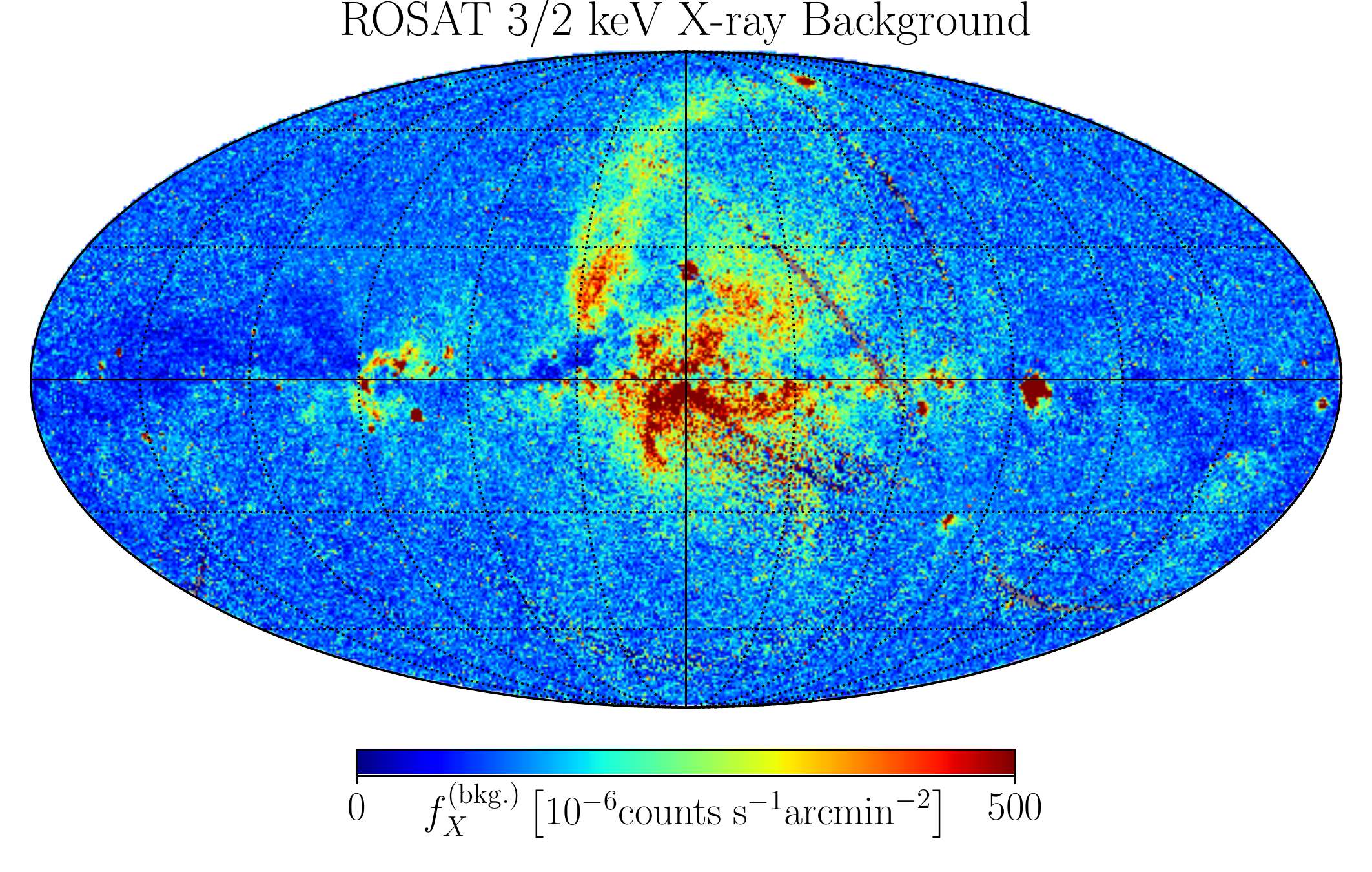}
    \caption{The ROSAT 1/4 (upper), 3/4 (middle) and 3/2 (bottom)~keV X-ray background map reproduced from Ref.~\cite{snowden1997} with same angular resolution ($\sim 0.2^{\circ}$) and unit of flux ($10^{-6} \text{counts s}^{-1}\text{ arcmin}^{-2}$) as provided in the data.}
    \label{fig:sxrb}
\end{figure}

\begin{figure*}[!htb!]
    \begin{center}
    \includegraphics[width=0.45\textwidth]{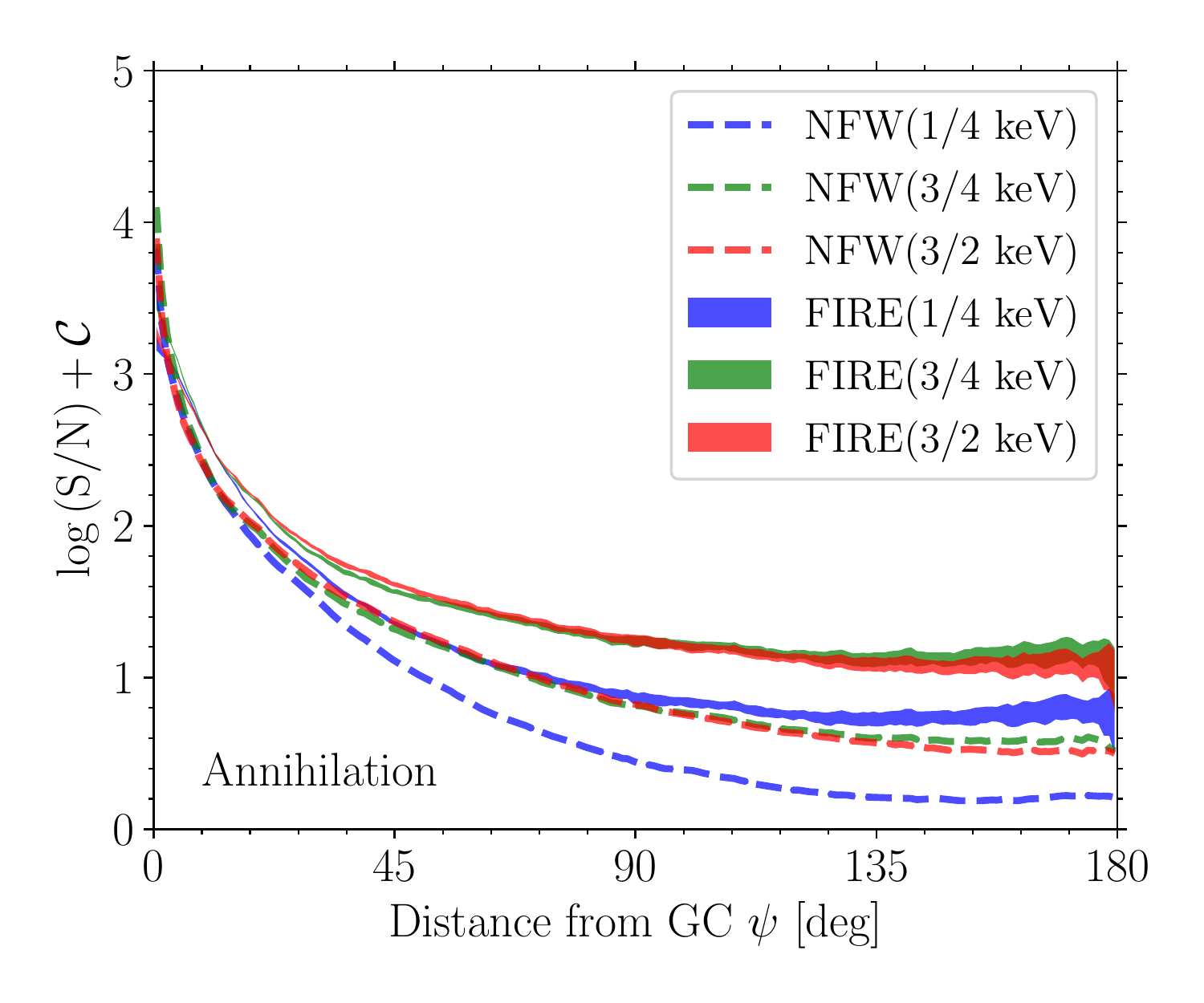}
    \hspace{10pt} \includegraphics[width=0.45\textwidth]{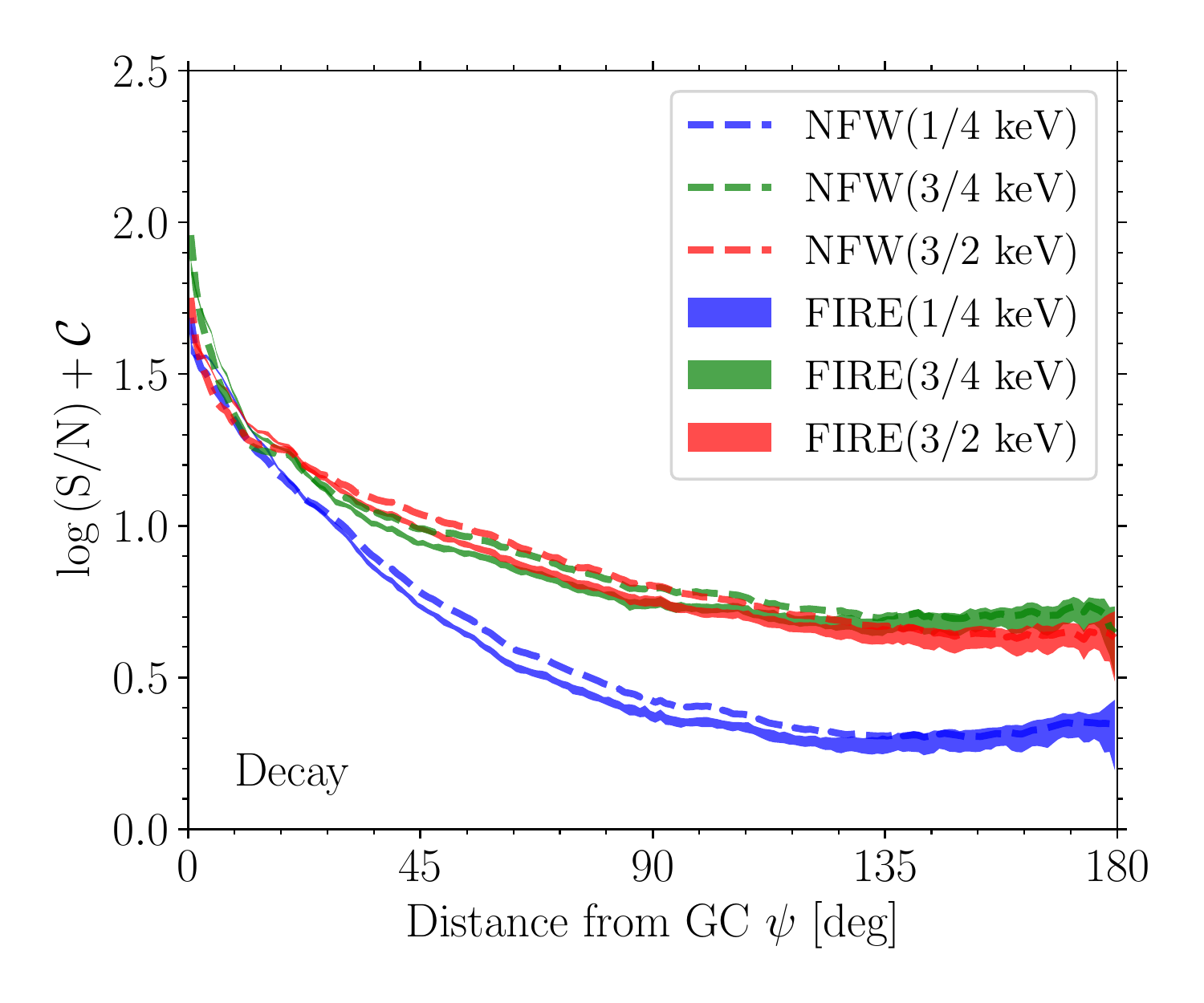}
    \end{center}
    \caption{Left: The average S/N of X-ray DM annihilation signal versus angle $\psi$ to the Galactic Center. The blue, green and red envelopes represent the FIRE-2 result against 1/4, 3/4 and 3/2~keV background. The envelopes are obtained by different luminosity at 16 observer positions in \textit{m12i} galaxy. As comparison, the blue, green and red dashed lines denote the Milky Way NFW results for 1/4, 3/4 and 3/2~keV background. The FIRE simulation's S/N is enhanced due to the substructure boost. Right: The variation of average relative S/N of X-ray DM decay signal versus angle $\psi$ to the Galactic Center. As expected, Galactic Center region has the most significant DM annihilation and decay signal against the background.  }
    \label{fig:xsncurve}
\end{figure*}

\begin{figure}[!htb!]
    \begin{center}
    \includegraphics[width=0.45\textwidth]{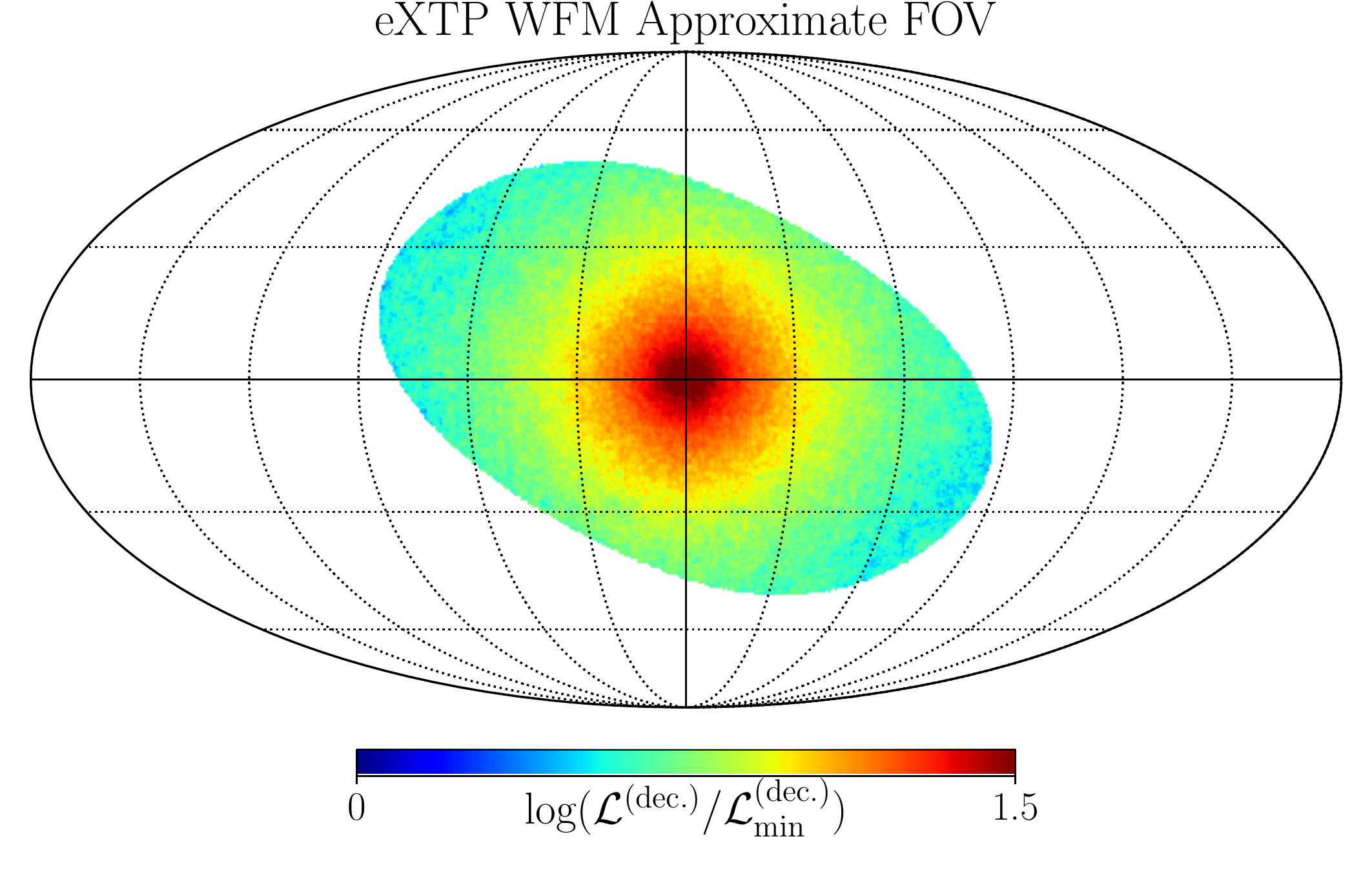}
    \end{center}
    \caption{The simulated eXTP WFM FOV approximated over the FIRE-2 DM decay luminosity map, matching the approximate FOV given in Fig.~4 of Ref.~\cite{Hernanz:2018llr}. The semi-major and semi-minor axis of the elliptical region are $90^{\circ}$ and $45^{\circ}$, respectively, which is an approximation of WFM's 3 Pairs FOV ($180^{\circ}\times 90^{\circ}$). Note that the FOV covers the most luminous Galactic Center region and would offer an unprecedented signal for Galactic DM decay within its band of 2-20 keV. Specifically, a 100~ks exposure of the region will collect between $\sim 10^5$ to $10^6$ events from the candidate 3.5~keV line.}
    \label{fig:eXTP}
\end{figure}

Let us now move to the analysis of the DM luminosity $\mathcal{L}_{\chi}$ (namely DM $D-$ and $J-$factors) estimated from FIRE-2 simulations.
Our goal in this section is twofold: \textit{i)} reassess the signal-to-noise ratio of DM annihilation/decay over realistic X-ray and $\gamma$-ray background according to state-of-the-art DM N-body simulations, that include effects from baryonic physics and also from subhalos; \textit{ii)} highlight the most important opportunities in the forthcoming observational campaigns in the X-ray and $\gamma$-ray band for DM indirect searches. 

Going back to Eq.~\eqref{eq:flux_DM} and leaving aside the spectral-shift effects discussed in the previous section, for an observer placed at the Sun position in the Galaxy ($R_{\odot} = 8.2$~kpc), the 
luminosity from the decay (annihilation) of DM particles in the Galactic halo is given by:
\begin{equation} 
\label{eq:DM_L}
    \mathcal{L}^{\textrm{dec.}(\textrm{ann.})}_{\chi} = \int_{\text{los}} \rho_{\chi}^{(2)}[r(\vec{s}\,)]d\vec{s} \ .
\end{equation}
Note that the resulting DM luminosity towards the GC is expected to be enhanced in the scenario of particle annihilation with respect to the DM decay one: in the former case it is sensitive to the DM number density squared. Moreover, the l.o.s. integral above crucially depends on the FOV chosen: a simple consideration that will be of extreme relevance in establishing the best observational prospects for DM astronomy in the X-ray band.

Following from what already elaborated on FIRE-2 simulations at the beginning of the previous section, the extraction of $\mathcal{L}_{\chi}$ from the \textit{m12i} MW-like galaxy essentially depends on the characterization of the distribution of DM particles in the simulation. On the basis of a total of $\sim 3\times 10^{6}$ DM particles with resolution $ M_{p} \simeq 3.5 \times 10^4 M_{\odot}$, for the analysis of the simulation output we design a suitable Cartesian grid and estimate $\rho_{\chi}(\vec{x})$ in each of the cubic cells of length $\Delta x$, counting the number of particles $N^{(\Delta x)}_{\, p}$ inside the cell. Hence, the estimate of the DM density in the cells follows the basic prescription $\rho_{\chi}(\vec{x}) \simeq N^{(\Delta x)}_{\, p} M_{p}/\Delta x^{3}$. Based on the approximate location of the Large and Small Magellanic Cloud, roughly distant $\mathcal{O}(50)$~kpc from the GC, we decided to adopt two different grids in order to optimize the computationally intensive evaluation of $\rho_{\chi}$: a fine sampling of the inner DM distribution set by $\Delta x = 0.1$ kpc and a less precise one for DM particles farther than 50~kpc from the GC using $\Delta x = 0.6$~kpc. Note that at such a distance from the GC, the DM density has already dropped significantly: this justifies a looser sampling without affecting the final outcome of our computation. Eventually, after evaluating the DM density in each of the cells composing a 300~kpc-sized cubic volume, we performed a linear interpolation to reconstruct the DM density profile $\rho(\vec{x})$ of the FIRE-2 \textit{m12i} MW realization. 

We test the agreement of the DM density distribution in the FIRE-2 MW-like simulation by comparing with the stellar kinematics constraints that probe the DM content of the Galaxy. We observe a good match with the density distribution of Refs.~\cite{Catena:2009mf,MW_NFW,2019JCAP...09..046K}. 
More specifically, taking an NFW profile as a reference halo, within a spherical average of the outcome extracted from the FIRE-2 simulation, a very good description of our DM profile at radii $\gtrsim$ 1~kpc is given by $\rho_{H} = 0.4$~GeV cm$^{-3}$ and $r_{H} = 16$ kpc\footnote{Note that due to the presence of baryonic physics in FIRE-2, departures from NFW behavior in the inner slope of the DM distribution were expected and seen in the very inner halo profile.}, already introduced in section~\ref{sec:level2} for the computation of the predictions reported in Table~\ref{tab:xraymission}.

The knowledge of $\rho(\vec{x})$ allows us to proceed in the evaluation of Eq.~\eqref{eq:DM_L}, i.e. the DM luminosity from FIRE-2. We compute all the 786432 l.o.s. in the skymap with HEALPix resolution of index~8, which has an angular resolution in the sky of about $0.23^{\circ}$ that is sufficient for our predictions. The luminosity map for the case of DM annihilation is displayed in the left panel of Fig.~\ref{fig:lumin_map} under Mollweide projection. In a similar fashion, we computed from FIRE-2 simulation \textit{m12i} the full-sky map for the DM decay scenario as well, illustrated in the right panel of the same figure. 

A first look at the maps in Fig.~\ref{fig:lumin_map} clearly reproduces the widespread expectation that the GC is by far the brightest spot of interest for DM. In an ideal world where background and foreground contamination would be completely absent, any other region of interest with a FOV not including the GC would result to be of much less relevance for the program of DM indirect searches. However, note that the granular structure present in both maps underlies the presence of substructures that populated the specific MW zoom-in simulation analyzed. In light of a more careful inspection of what would be the signal-to-noise ratio expected in the X-ray and $\gamma$-ray band, the role of subhalos may be crucial in enhancing a DM signal along the l.o.s. as well as confirming or disproving any possible information on DM extracted from the analysis of the GC region itself.

With the final aim of providing a valuable insight on the signal-to-noise ratio (S/N) for a DM discovery, let us start with a brief discussion on the expected signal-to-noise ratio from DM annihilation/decay in the X-ray band. 
While the sensitivity of current observational X-ray missions in the quest for DM is limited by their particle instrumental errors, the main source of noise for a fair S/N estimate carried out by future X-ray campaigns may be provided by the X-ray background, whose origin constitutes a broad field of research~\cite{1992ARA&A..30..429F,Hickox:2005dz}. At energies roughly above few~keV, the X-ray sky is expected to be dominated by extra-galactic contributions, see for instance the study of SuperAGILE~\cite{2010A&A...510A...9F} and the more recent measurements from Swift-BAT~\cite{Oh:2018wzc}. On the other hand, the soft diffuse X-ray background features the complexity of Galactic physics in the turbulent ISM, and it was measured more than 20 years ago by ROSAT~\cite{snowden95,snowden1997}. The latter still provides today the most precise observations of cosmic photons in the energy range 0.5~-~2~keV, and the only accurate full-sky maps in the broad X-ray band that are not affected by small exposure times, typically implied for the hard X-ray spectrum in relation to point-source detection.

Consequently, we opted for the use of ROSAT maps as a good proxy of the noise expected in the soft X-ray background, leaving the inspection of the hard X-ray sky in our S/N analysis to future investigation. 
In order to process ROSAT dataset, we used the online tools described by the collaboration in Ref.~\cite{sxrbg}\footnote{\href{https://heasarc.gsfc.nasa.gov/cgi-bin/Tools/xraybg/xraybg.pl}{https://heasarc.gsfc.nasa.gov/cgi-bin/Tools/xraybg/xraybg.pl}}; we obtained photon counts for a given l.o.s. and we estimated the flux according to the formula~\cite{sxrbg}:\footnote{For ROSAT data and the flux calculation in Fig.~\ref{fig:sxrb}, please refer to \href{https://heasarc.gsfc.nasa.gov/Tools/xraybg\_help.html\#references}{https://heasarc.gsfc.nasa.gov/Tools/xraybg\_help.html\#references}.} 
\begin{equation}
    f_{\rm X}^{\text{(bkg)}} = 0.39063 \, \left( \frac{\text{counts - background}}{\text{exposure}} \right) \text{arcmin}^{-2}\;\text{s}^{-1} \ .
\end{equation}
 In Fig.~\ref{fig:sxrb} we show the result of this procedure: using the resolution of the data measured by ROSAT, of about $0.2^{\circ}$, we obtained three different probes of the X-ray background at the energy $E = 1/4, 3/4, 3/2$~keV. Note that the angular resolution involved is consistent with the chosen resolution for the DM luminosity in Fig.~\ref{fig:lumin_map}.

Hence, we proceed constructing the corresponding full-sky noise maps (again with HEALPix resolution of index~8), exploiting the estimated X-ray luminosity for DM and the ROSAT maps at the three energies provided by the collaboration.
Assuming an exact Poisson statistics for the photon counts, the S/N can be easily forecast as the ratio between the expected DM photon counts, $N_{s}$, and the square root of the estimated background photon $\sqrt{N_{\rm bkg}}$. Therefore, up to dimensional factors that would depend on the characterization of the fundamental nature of DM, and on the specifics of an experiment, the S/N would be predicted generically  as
\begin{equation}
  \
   \text{S/N} \propto \mathcal{L}_{\chi} \Big/ \sqrt{f^{\text{(bkg.)}}} \ ,
\end{equation}
where $\mathcal{L}_{\chi}$ stands for the DM luminosity from the FIRE-2 simulation, while $f^{\text{(bkg.)}}$ is the background flux considered; in this case, it corresponds to $f_{\rm X}^{\text{(bkg.)}}$, namely the X-ray background flux measured by ROSAT in $10^{-6} \text{counts s}^{-1}\text{ arcmin}^{-2}$. 

Selecting 15 l.o.s. in the sky at varying angular distance from the GC, $\psi$, we ended up evaluating the average S/N as a function of $\psi$ and its typical spread, as reported in Fig.~\ref{fig:xsncurve}. As expected, from Fig.~\ref{fig:xsncurve}, regardless of the different morphological information contained in the three maps in energy for the inferred X-ray noise, the emerging S/N picture always favors DM detection in proximity of the Galactic center region. Going from $\psi = 180^{\circ}$ towards the GC, we observe a gain in S/N of roughly one order of magnitude (for 1/4 keV even more) for the scenario involving DM decay, and a jump of almost three orders of magnitude in S/N for the case of DM annihilation.
\begin{figure*}[!htb!]
    \begin{center}
    \includegraphics[width=3.4 in]{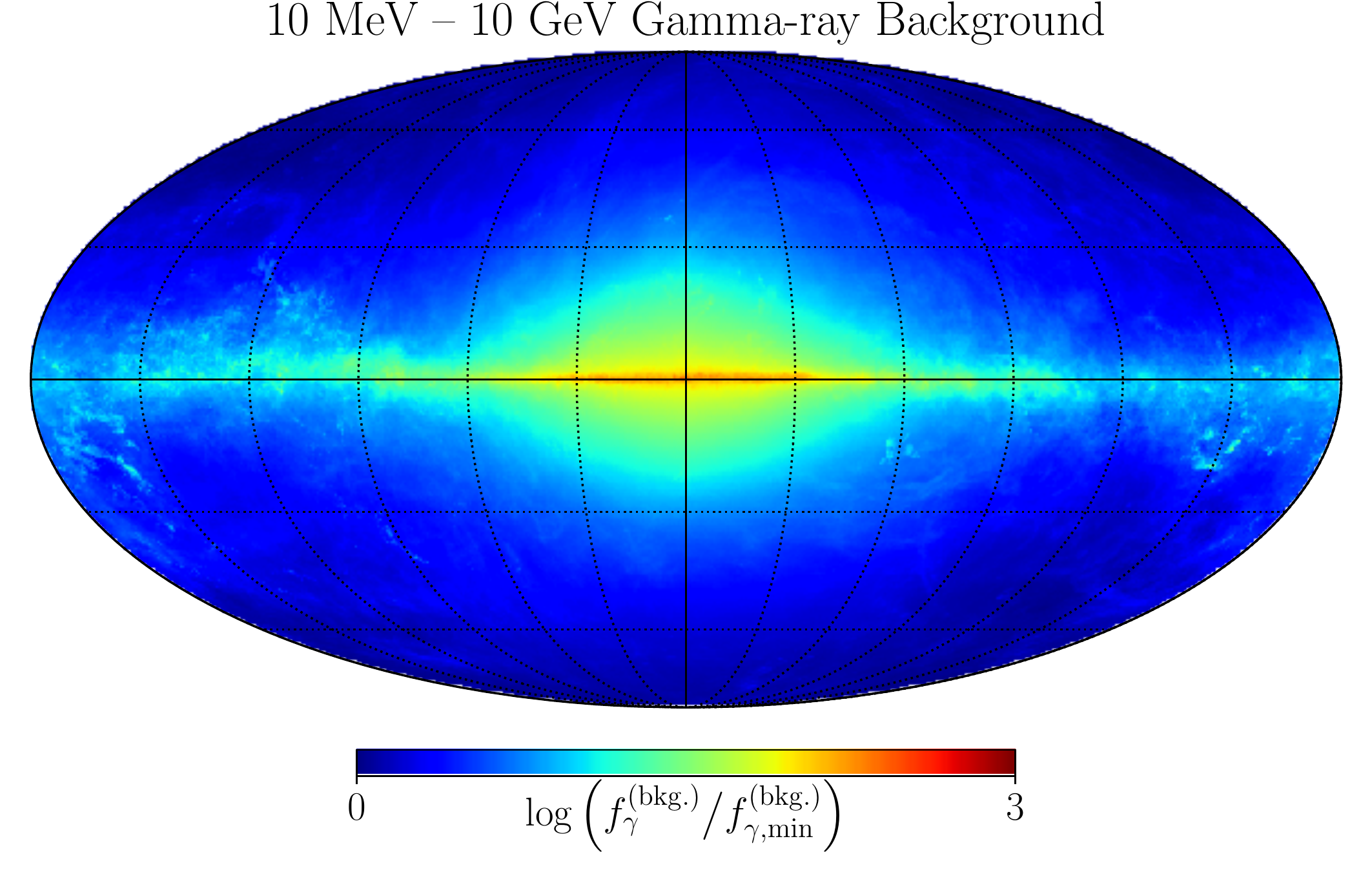}
    \hspace{10pt} 
    \includegraphics[width=3.4 in]{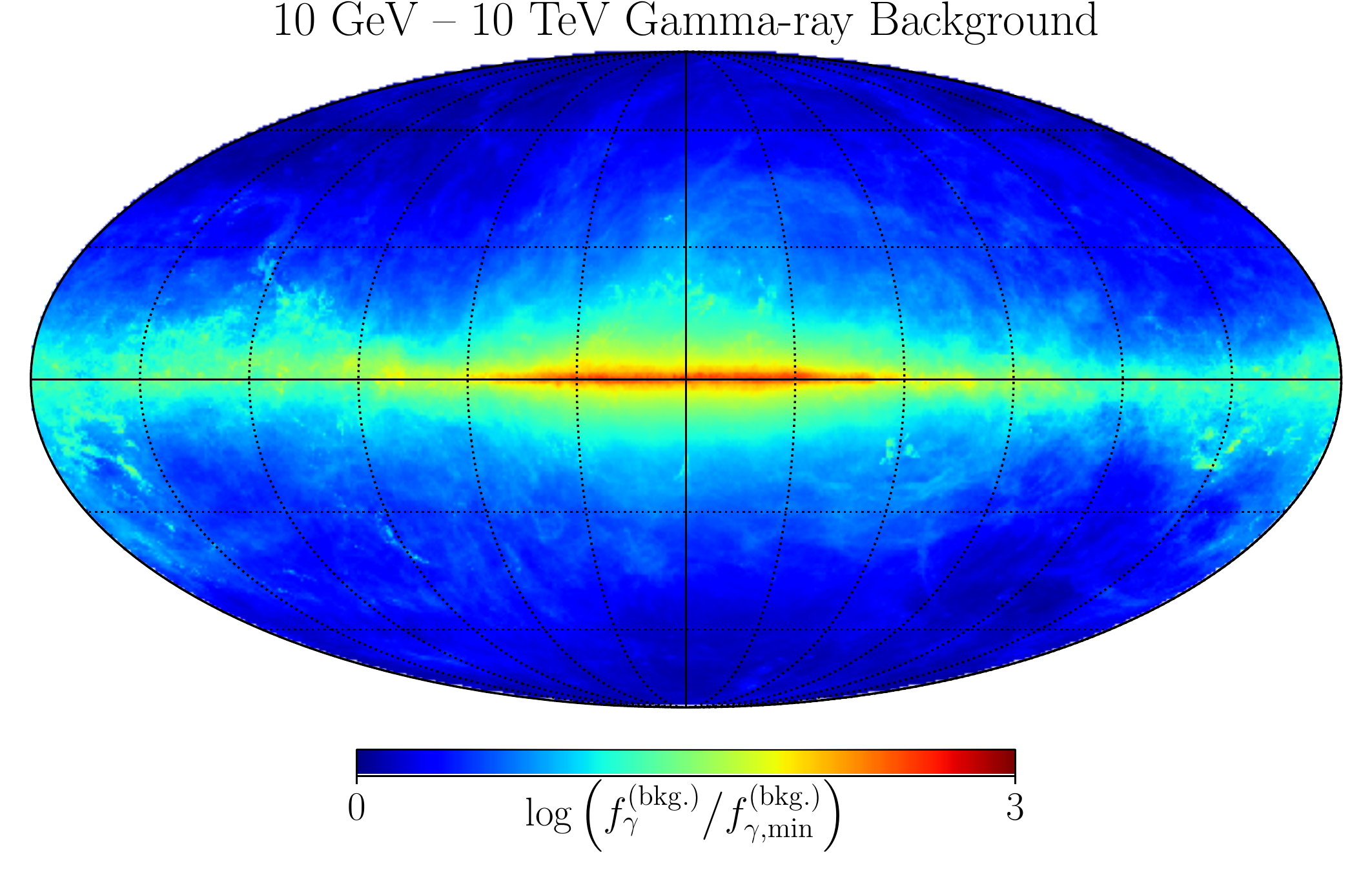}
    \end{center}
    \caption{Full-sky map for the gamma-ray background obtained according to the CR propagation model derived in Ref.~\cite{Fornieri:2019ddi}, fitting all local CR measurements, and simulated with the DRAGON code~\cite{Evoli:2016xgn,Evoli:2017vim}. Left panel shows the prediction related to the low-energy range of 10~MeV~-~10~GeV, while the right panel shows the prediction at high energies, 10~GeV~-~10~TeV.}
    \label{fig:gamma}

\end{figure*}

\begin{figure*}[!htb!]
    \begin{center}
    \includegraphics[width=3.4 in]{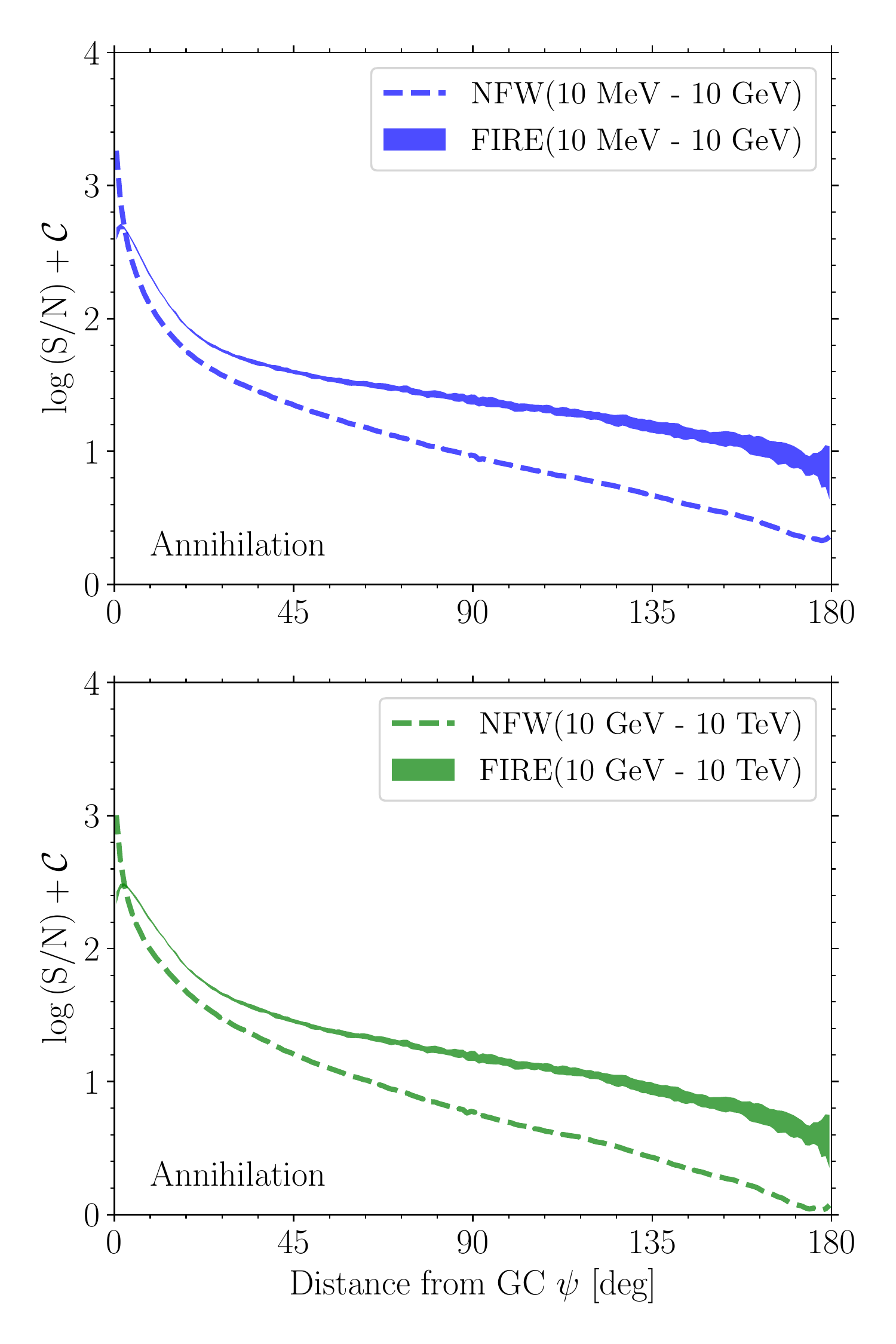}
    \hspace{10pt} \includegraphics[width=3.4 in]{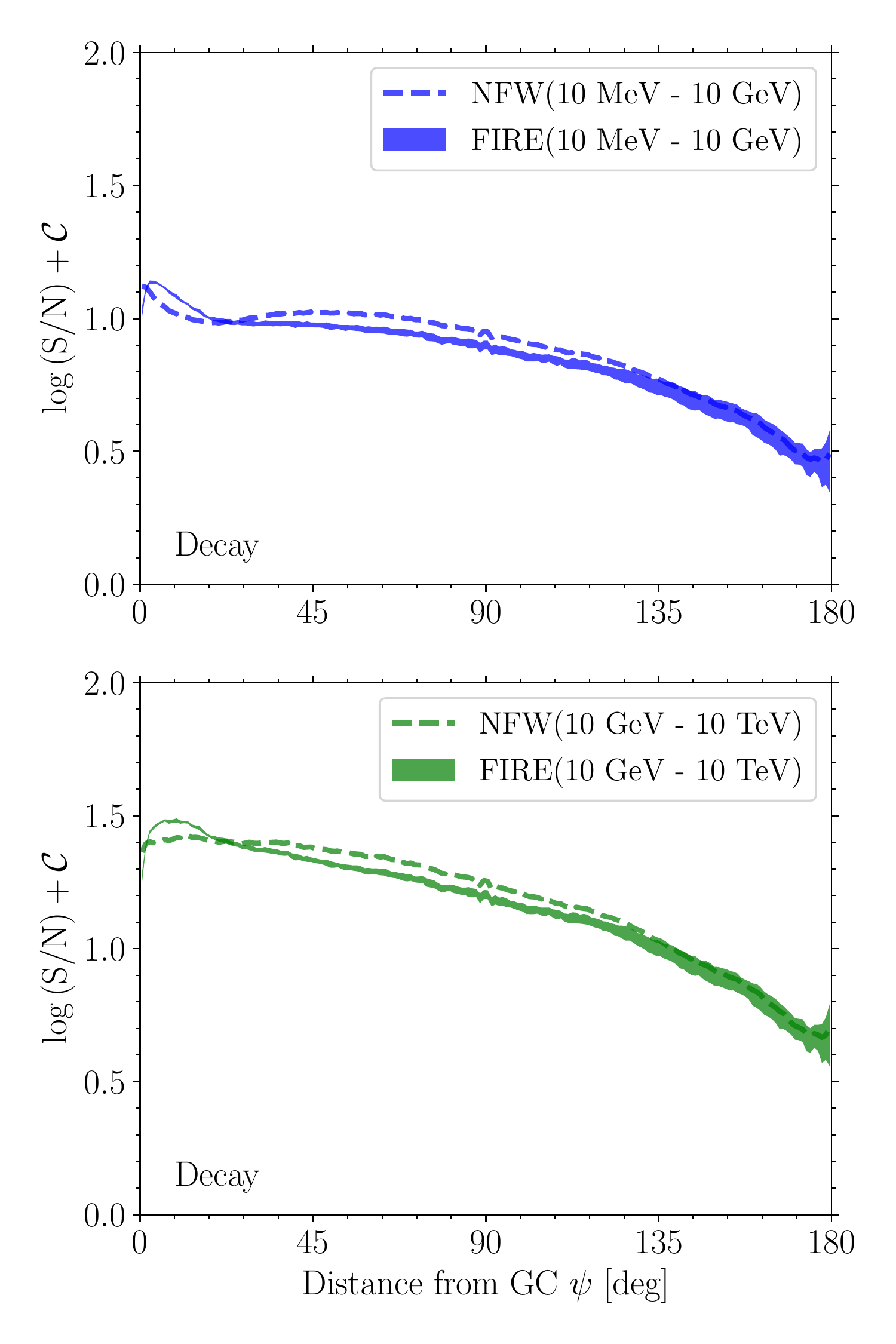} 
    \end{center}
    \caption{Left: The variation S/N of gamma-ray DM annihilation signal versus angle $\psi$ to the Galactic Center. The blue and green envelopes represent the FIRE-2 result against $10\text{ MeV} - 10\text{ GeV}$ and $10\text{ GeV} - 10\text{ TeV}$ simulated background. As comparison, the blue and green dashed lines denote that from a smooth Milky Way NFW halo model from a $10\text{ MeV} - 10\text{ GeV}$ and $10\text{ GeV} - 10\text{ TeV}$ simulated background. The FIRE simulation's S/N is enhanced due to the substructure boost. Right: The analogous variation of S/N for the case of gamma-ray DM decay versus angle $\psi$ to the Galactic Center. }
    
     \label{fig:gsncurve}
\end{figure*}

On the basis of the results obtained above, we would like to pause one moment on an important remark, i.e. what seems to us a highly promising forthcoming opportunity for DM searches in the X-ray wavelength. 
Overall, luminosity measures of DM emission dominate over the above-mentioned velocity measures for large FOV observations. Importantly, an observatory with a wide FOV can have very high sensitivity because of the immense DM mass within the field of view, increasing the signal considerably, while the background still scales as $\sim \sqrt{f^{\rm bkg.}}$. That is, the overall sensitivity increases as $S/N \sim \text{FOV}/ \sqrt\text{FOV}\sim \sqrt\text{FOV}$, in the approximation of a constant DM density in the FOV. As shown in previous work, the XQC sounding rocket with 100 s exposure of an X-ray microcalorimeter to a FOV of 0.81 sr toward the Milky Way Galactic Cap \cite{McCammon:2002gb} placed limits comparable to a 3 Ms exposure with {\it Chandra}'s 8 arcmin FOV in its deep field observations, when matching at the the same energy/mass range \cite{Abazajian:2006jc}. 

The WFM instrument \cite{Hernanz:2018llr} aboard the eXTP X-ray Telescope \cite{Zhang:2016ach} will have a field of view of $\approx\,$4 sr and an energy resolution comparable to {\it Chandra}. This will provide an unprecedented high-signal measure of DM photons. Given that each camera pair on the WFM aboard eXTP has an effective area  $A_{\text{eff}} \approx 79\text{ cm}^{2}$, and adopting the candidate particle mass and decay rate at $m_s = 7\text{ keV}$ and $\Gamma_{\chi} = 1.3\times 10^{-28} \text{ s}^{-1}$, we can approximate the signal count for WFM using both the simulations here, and analytic halo models, which agree to large extent given the broad FOV (Figure \ref{fig:eXTP}). Each of WFM's camera pairs' have a FOV of $30^\circ\times 30^\circ$ at full exposure, and a reduced exposure out to $90^\circ\times 90^\circ$ \cite{Hernanz:2018llr}. Our FIRE-2 simulation estimates a signal count of 3.5~keV sterile neutrino decay photons, in the case of the smallest FOV to be of $N_s\approx 390000$, for a relatively brief exposure of $T_{\text{expo}} = 100\,\text{ks}$. The larger $90^\circ\times 90^\circ$ gives $N_s\sim 1.3\times 10^6$. This signal rate is consistent with smooth halo models with our canonical NFW parameters in Section~\ref{sec:level2}. 

The WFM instrument's particle instrumental background and astrophysical X-ray background will need to be analyzed to reveal the exact sensitivity of WFM, though the signal is large. We estimate the signal relative to the X-ray background given in Gruber et al.~\cite{Gruber:1999yr}, which is an overestimate of the expected X-ray background as WFM will resolve bright point sources. We set aside WFM instrumental backgrounds for future work.  Taking the Gruber et al.\ cosmic X-ray background at 600 eV around the 3.5 keV line (twice WFM's FWHM), for the case of a $30^\circ\times 30^\circ$ full exposure signal toward the GC, the signal-to-noise ratio is $\sim\! 170$, and the signal-to-noise for $90^\circ\times 90^\circ$ is $\sim\! 270$. Therefore, with respect to the cosmic X-ray background, we estimate that WFM is very sensitive to the candidate 3.5 keV line due to its large FOV toward the GC. Importantly, note that a 3.5 keV atomic transition from the X-ray emission of the MW warm-hot halo ($\sim$200 eV) will have a Boltzmann-suppressed emissivity by a factor of $\sim\,$20000 relative to local hot plasma in the MW or a cluster of galaxies at $\sim 2\ \rm keV$. Overall, we find WFM will likely be more sensitive than the other instruments aboard eXTP, whose FOV are $\sim\! 10$ arcmin \cite{Malyshev:2020hcc}.  

Let us finally end our discussion with the estimate of the S/N in the gamma-ray band. Indeed, future ground-based experiments like CTA will study the GC region with unprecedented sensitivity~\cite{Bigongiari:2016amk, Acharya:2017ttl}, and will probe an energy scale well beyond the TeV, while space missions like GAMMA-400 and, in particular, e-ASTROGAM may give us new insights even at lower energies, approximately down to the MeV scale~\cite{Egorov:2017nyt,eastrogam}.

According to these considerations, we have organized our prediction for the gamma-ray background into two bins: $10\text{ MeV} - 10\text{ GeV}$ and $10\text{ GeV} - 10\text{ TeV}$ photons. 
In this two energy ranges, the full-sky emission detected by experiments like Fermi-LAT~\cite{ackermann12} proved us in great detail that more than 80\% of the collected photons are correlated with the Galactic plane and with gas and starlight distributions. This fact further lead us to the relevance of Galactic CR physics behind the scenes of the observed gamma-ray sky. 
In particular, high-energy cosmic rays of Galactic origin, interact with the magnetized ISM and with the interstellar radiation field (ISRF) to yield three fundamental contributions: photons produced mainly by neutral meson decay, the so-called $\pi^{0}$ component, sensitive to the hydrogen gas distribution of the ISM; the bremsstrahlung contribution due to the relativistic charged CRs traveling through the ISM, again proportional to the ISM gas density; the inverse-Compton (IC) component, related to the kick in energy that CMB photons and starlight receive from high-energy leptons accelerated in the Galaxy. 

In order to study the gamma-ray noise for an accurate S/N full-sky prediction, here we exploited the outcome of the recent analysis carried out in Ref.~\cite{Fornieri:2019ddi}, that presented a new benchmark for CR propagation in the Galaxy: The authors constrained CR propagation fitting to all the up-to-date measurements relative to local CR observables. We therefore reproduced the Galactic CR simulation of Ref.~\cite{Fornieri:2019ddi} using the numerical package DRAGON~\cite{Evoli:2016xgn,Evoli:2017vim}. Then, adopting the standard assumption that CR properties probed locally may hold effectively across the whole Galaxy, we computed the gamma-ray emissivities associated to the $\pi_{0}$, IC, and bremsstrahlung components. For this last step, we adopted state-of-the-art results for what concerns the ISM gas distribution~\cite{2015A&A...582A.121R} and the modelling of the ISRF~\cite{Vernetto:2016alq}. 
Our full-sky prediction for the expected gamma-ray background is reported in Fig.~\ref{fig:gamma}, obtained once again with HEALPix resolution index of~8. In this novel prediction, interestingly we note that the $10\text{ MeV} - 10\text{ GeV}$ sky presents a relatively important contribution coming from leptonic diffusion and its convolution with the ISRF distribution, while the hadronic component greatly dominates the scene of the gamma-ray sky at energies above $10~\text{GeV}$.

With our prediction of the gamma-ray background, we look at the computation of the S/N for DM annihilation and decay in this wavelength as a function of the angular distance from the GC, $\psi$. As shown in Fig.~\ref{fig:gsncurve}, the GC region stands out also in this case as the most compelling direction for the discovery of DM. In particular, we observe that the impact of baryonic physics is important at very small $\psi$. There remains two orders of magnitude of enhancement of the signal-to-noise as  $\psi$ approaches the GC region ($\psi\rightarrow 0^{\circ}$). Therefore, the GC remains a primary target for all the next-generation gamma-ray campaigns whose science case includes the quest for particle DM. Note, however, that the GC region ($\psi \lesssim 20^\circ$) also remains one of the most challenging regions for which our adopted foreground models are insufficient. See, e.g., ~\cite{Gaggero:2015nsa, Carlson:2016iis}.

In this regard, a very non-trivial outcome is captured both in Figs.~\ref{fig:xsncurve}~and~~\ref{fig:gsncurve} when one compares the decay and annihilation scenarios and focuses the attention also to the S/N obtained for the case of the analytic NFW profile reported.
As anticipated before, except for the very inner kpc from the GC, the NFW profile shown turns out to be a very good proxy of the smooth DM halo component of the MW-like galaxy analyzed in the simulation: therefore, it gives us a glimpse of the role of substructures along the line-of-sight. In both Figs.~\ref{fig:xsncurve}~and~\ref{fig:gsncurve}, the S/N extracted from the FIRE-2 analysis agrees well with the NFW one for the case of DM decay. The NFW profile does not agree well for the annihilation scenario, which shows a departure of almost an order of magnitude between the two cases. This suggests us that the subhalo contribution along the l.o.s. in the FIRE-2 becomes relevant for the study case of DM annihilation. The presence of these subhalos is also significant in the S/N full-sky map in the X-ray and $\gamma$-ray bands, illustrated in Figs.~\ref{fig:xsnmap}~-~\ref{fig:gsnmap}. Therefore, while the GC region should be explored in great detail by several X-ray and gamma-ray surveys, for the specific case of DM annihilation,  bright substructures along the l.o.s. remain an important signal source in the potential upcoming era of DM astronomy.

\section{\label{sec:level4}Conclusion}
Future X-ray and gamma-ray missions will largely extend the capability of astronomical searches for DM via photon emission from annihilation, decay or internal structure, and potentially lead us to the era of ``Dark Matter Astronomy." We focus on near to long-term prospects for X-ray and gamma-ray observatories. We go beyond prior studies of the luminosity profile of DM annihilation or decay by using detailed hydrodynamic simulations of our MW Galaxy's formation with a representative DM spatial and velocity distribution, via the FIRE-2 Latte simulations. Narrow line features in energy are ubiquitous at higher order in DM emission (e.g., \cite{Abazajian:2001vt,Abazajian:2011tk}), which allows DM astronomy to involve spectroscopic studies revealing bulk velocity as well as dispersion motion of the DM. We have studied how the bulk velocity, velocity dispersion, and luminosity profile of DM signals will present themselves in the sky, using the FIRE-2 Latte suite of simulations of the Local Group. 

For bulk velocity spectroscopy, astrophysical foregrounds will be persistent in any future signal studies. For the case of narrow photon line found in the MW, there are definitive ways of differentiating a DM line from an astrophysical line given the Doppler velocity information. Ref. \cite{speckhard16} showed that telescopes with energy resolution $\mathcal{O}(0.1\%)$ are suitable for differentiating DM via a particle physics model-independent property of DM: the motion of DM inside the MW relative to baryonic gas. 

We employed a FIRE-2 cosmological hydrodynamic simulation from the Latte suite in order to more accurately specify the requirements of velocity spectroscopy to differentiate a DM and astrophysical line source. A pair of symmetric observations equally separated about  $l= 0^{\circ}$ to compare  the location of two detected emission lines' centroids could provide sufficient information for the DM diagnosis. We find that $l=\pm90^{\circ}$ would be an ideal place to make use of the velocity energy shift, giving the most significant separation of a DM line relative to an astrophysical source. For the representative case of the 3.5~keV candidate line, DM velocity spectroscopy performed by the forthcoming X-ray telescopes XRISM, Athena and Lynx are only capable of separating a DM line from astrophysical foreground with fairly long exposures. We estimate that the separation of the lines at $5\sigma$ would require an exposure of $\approx\! 4$ Ms to $\approx\! 15$ Ms of XRISM at $l=\pm90^{\circ}$, away from the Galactic Plane, with the range depending on the level of the foreground/background astrophysical emission. Note that this velocity procedure is applicable for any DM narrow features from annihilation or decay at X-ray or gamma-ray band. Overall,  sensitivity to Doppler velocity spectroscopy in the X-ray will require Athena and Lynx energy and flux sensitivity levels. DM overdensities in collapsed structures outside our MW halo remain a promising prospect for the narrow FOV observatories \cite{Abazajian:2001vt,Aharonian:2016gzq}. Velocity broadening in the X-ray and velocity spectroscopy in the gamma-ray will require mission sensitivities beyond those currently considered.

We studied the signal-to-noise of DM annihilation and decay in full-sky observations using optimal available foreground models. Most significantly, we point out that the luminosity of DM on the full sky remains a very robust signal when combined with large FOV observatories. Specifically, we find the WFM instrument aboard the eXTP Telescope to be a very well-suited large FOV instrument to search for DM signals on the sky, as well as follow up the 3.5~keV candidate signal. We estimate that WFM aboard eXTP would have a very high flux of the 3.5 keV candidate line, with a 100 ks exposure giving DM decay event counts between $10^5$ and $10^6$ depending on the acceptance across the field of view, with a commensurate high signal-to-noise, $\rm S/N\gtrsim 180$, relative to the cosmic X-ray background. 

In a more general sense, we also estimate the all-sky DM luminosity and the S/N curves from the X-ray to gamma-ray bands. Although the Galactic Center region has luminous foreground emission at X-ray or gamma-ray band, the most significant DM signals could be also found in this place. From the comparison of NFW profile for the MW, we conclude that the subhalos would have significant contribution on the annihilation emission in searches in outer regions of the halo.

Overall, the near term prospects for DM searches in high-energy photons remain promising. As X-ray and gamma-ray observatories improve, finer energy and spatial resolution have the potential to reveal not only spatial structure but also dynamical, velocity structure of the DM in our MW's halo, both in its bulk motion as well as its intrinsic velocity broadening. Recent candidate detections of anomalies in gamma-rays and X-rays may be the start of an era of DM astronomy, or novel signals across the electromagnetic spectrum could reveal surprises that can be fully explored by new observatories, guided by robust simulations of DM in our MW galaxy.

\begin{acknowledgments}
We thank Sijie Yu for suggestions about the usage of the Latte simulation; we acknowledge John Beacom, Federico Bianchini, Daniele Gaggero, and Dan Hooper for relevant discussion. We are also grateful to the authors of Ref.~\cite{Fornieri:2019ddi} for providing us the output of their wide-ranging analysis, and we thank Margarita Hernanz, Denys Malyshev, Andrea Santangelo, Jean in 't Zand, and Shuang-Nan Zhang for providing guidance with regards to eXTP's WFM technical capabilities.  D.Z. thanks Peilin Pan for support. K.N.A. and M.V. are supported, in part, by the National Science Foundation under Grants No.~PHY-1620638 and No.~PHY-1915005. 
\end{acknowledgments}

\appendix

\begin{figure*}
    \begin{center}
    \includegraphics[width=0.45\textwidth]{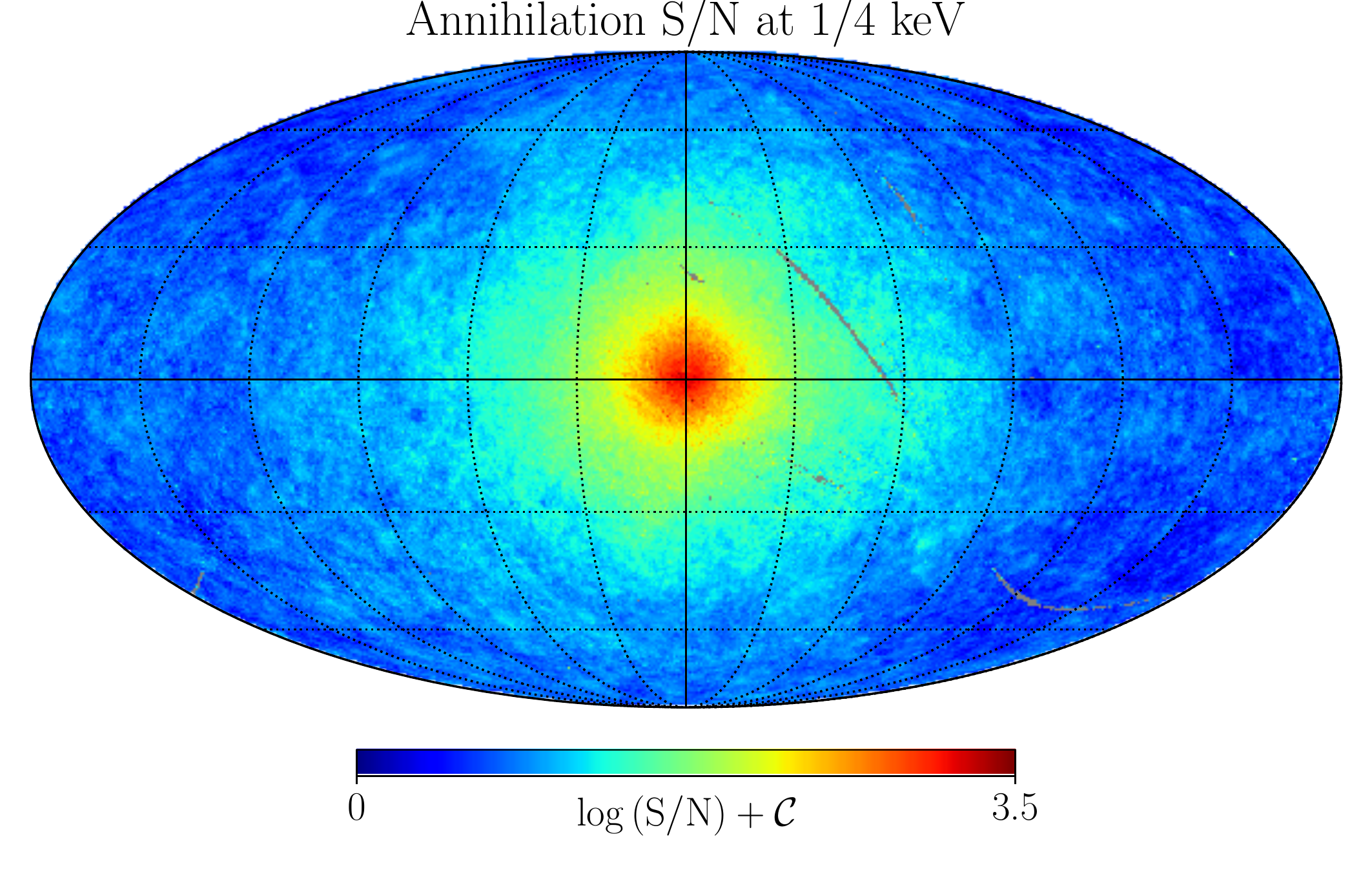}
    \hspace{10pt} \includegraphics[width=0.45\textwidth]{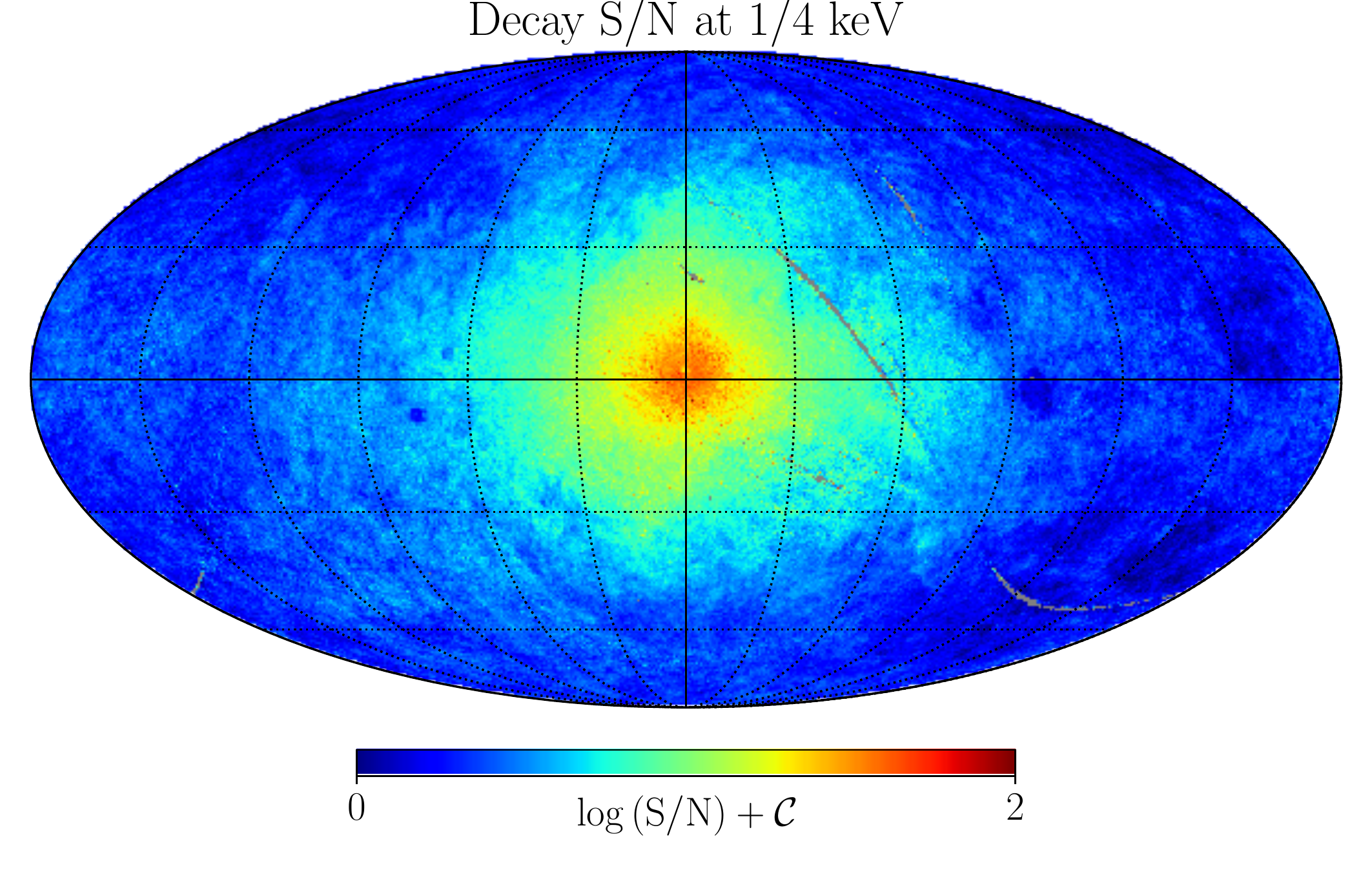} \\ \vspace{3pt}
    \includegraphics[width=0.45\textwidth]{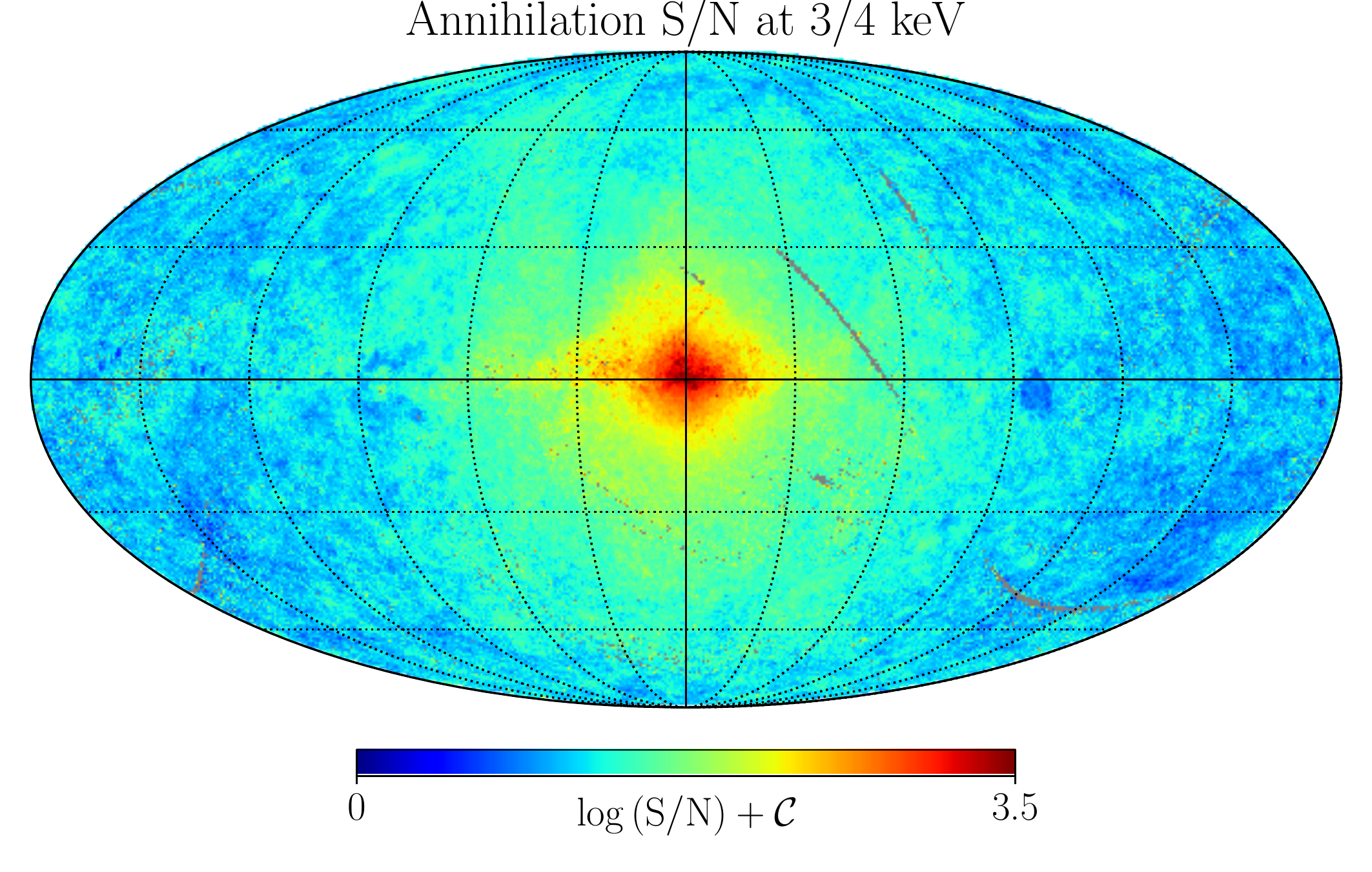}
    \hspace{10pt} \includegraphics[width=0.45\textwidth]{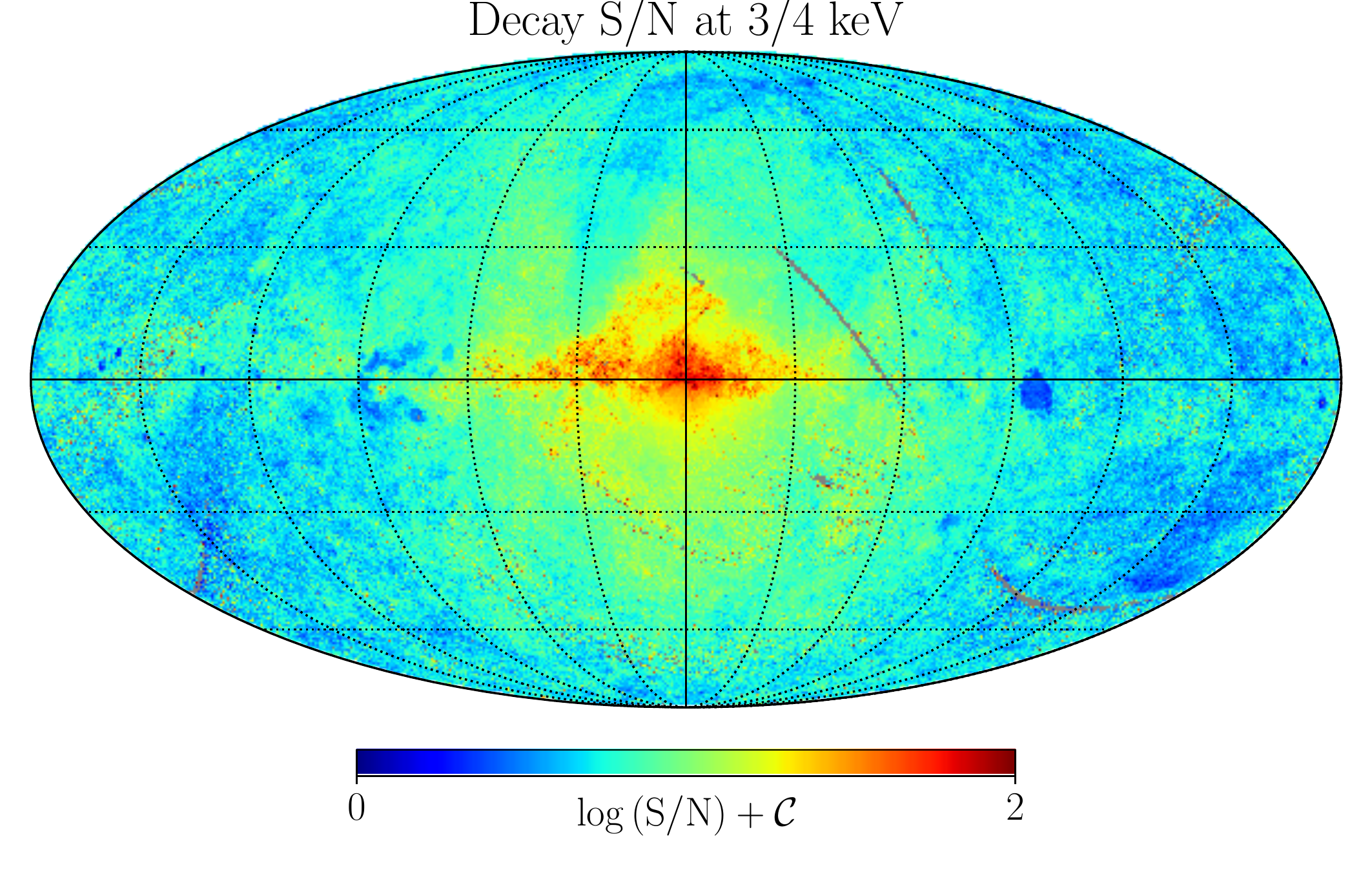} \\ \vspace{3pt}
    \includegraphics[width=0.45\textwidth]{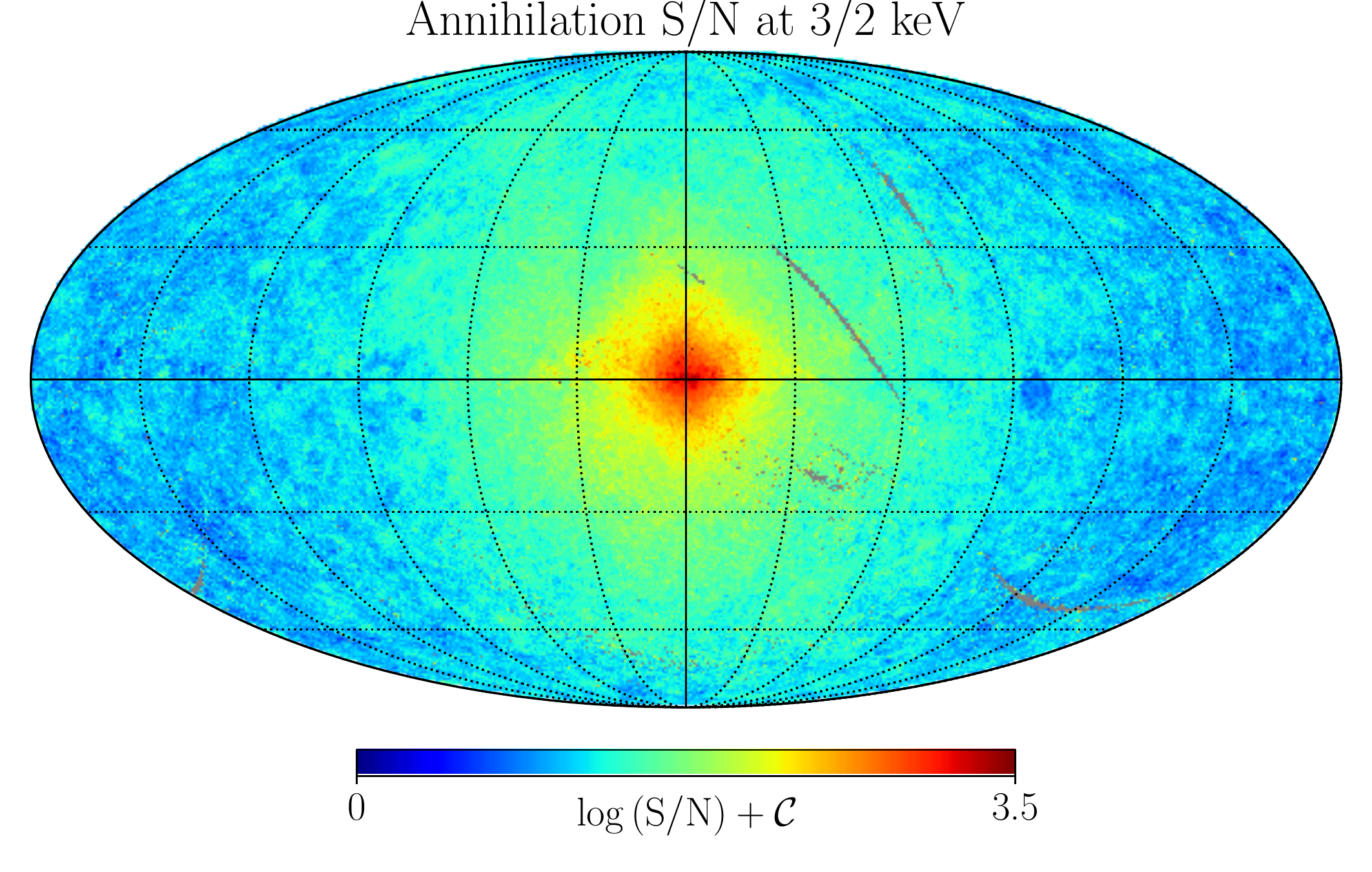}
    \hspace{10pt} \includegraphics[width=0.45\textwidth]{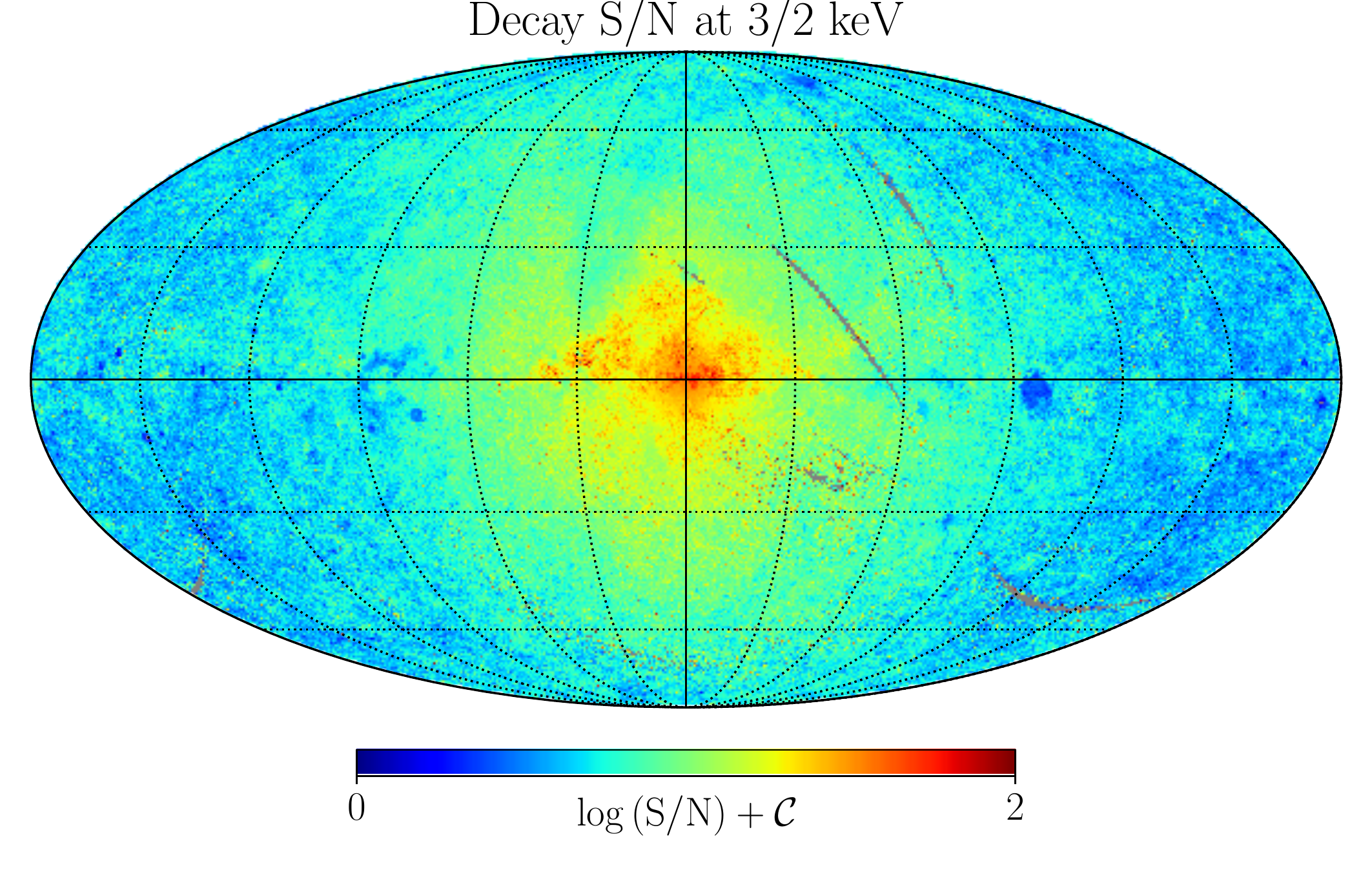}
    \end{center}
    \caption{The all-sky S/N prediction of DM X-ray signal at 1/4 (top), 3/4 (middle) and 3/2 (bottom) keV scale for annihilation (left) and decay (right) emission. The grey feature in the maps is due to the lack of soft X-ray background data.}
    \label{fig:xsnmap}
\end{figure*}

\begin{figure*}
    \begin{center}
    \includegraphics[width=0.45\textwidth]{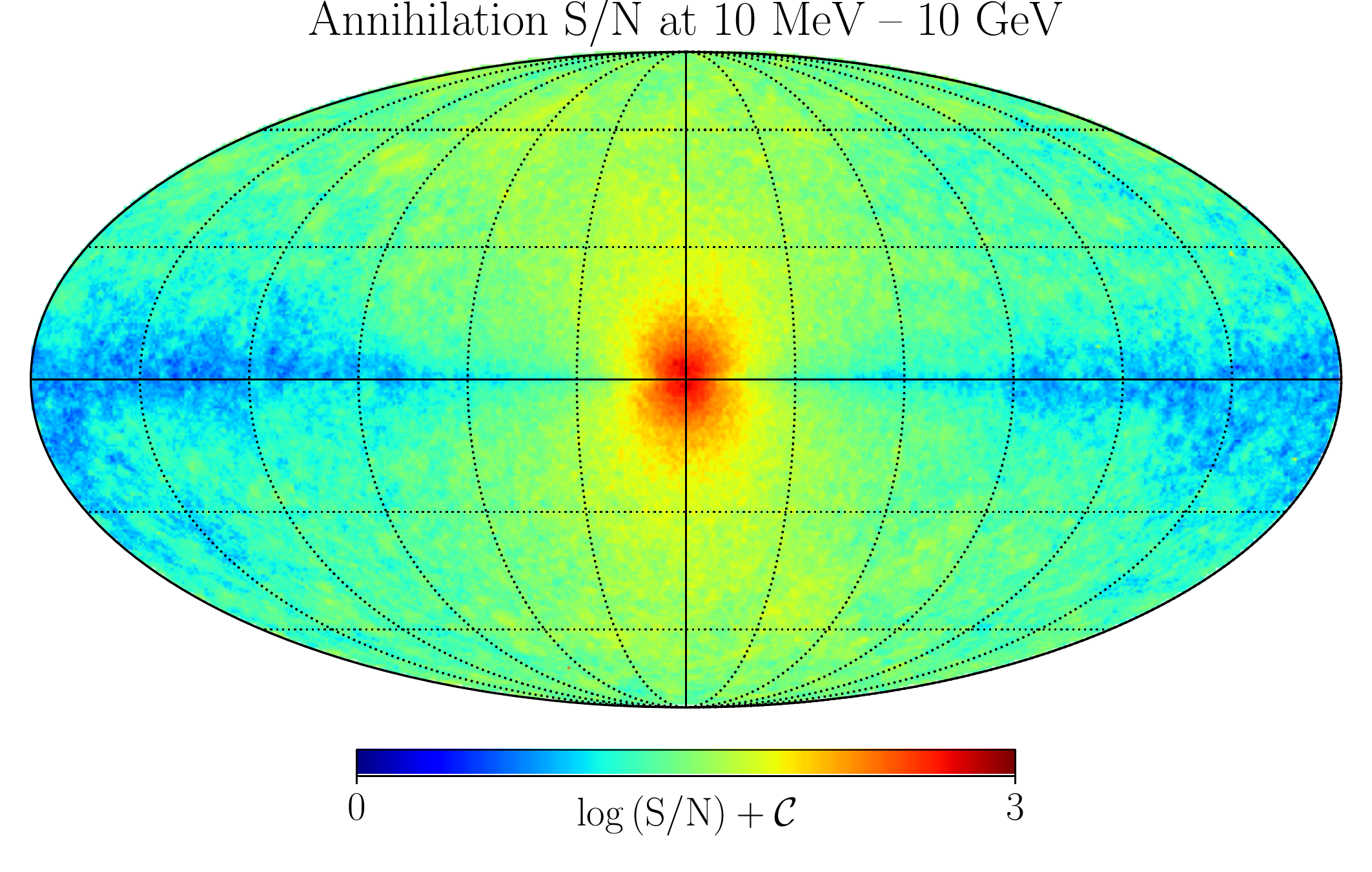}
    \hspace{10pt} \includegraphics[width=0.45\textwidth]{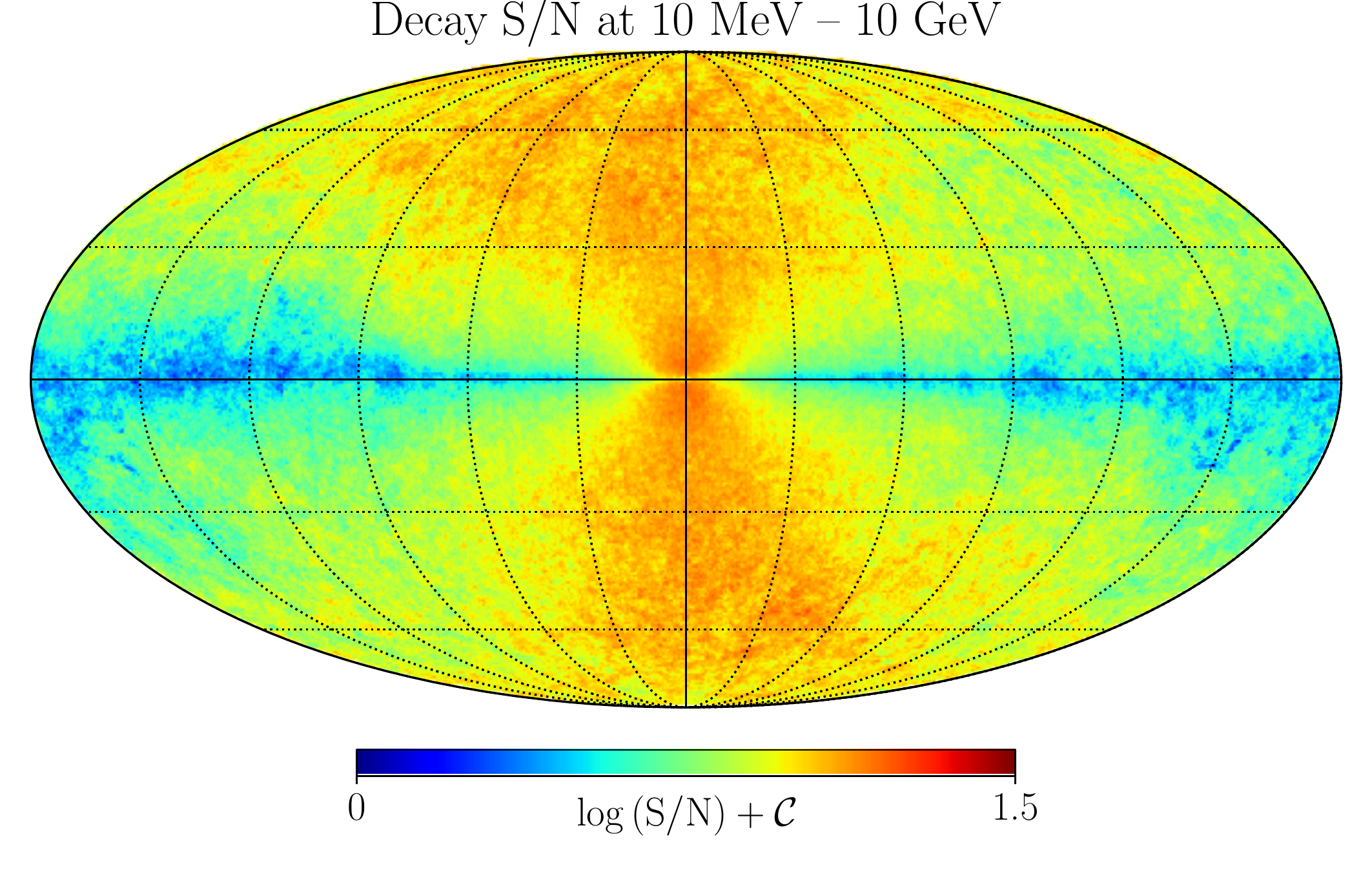} \\ \vspace{3pt}
    \includegraphics[width=0.45\textwidth]{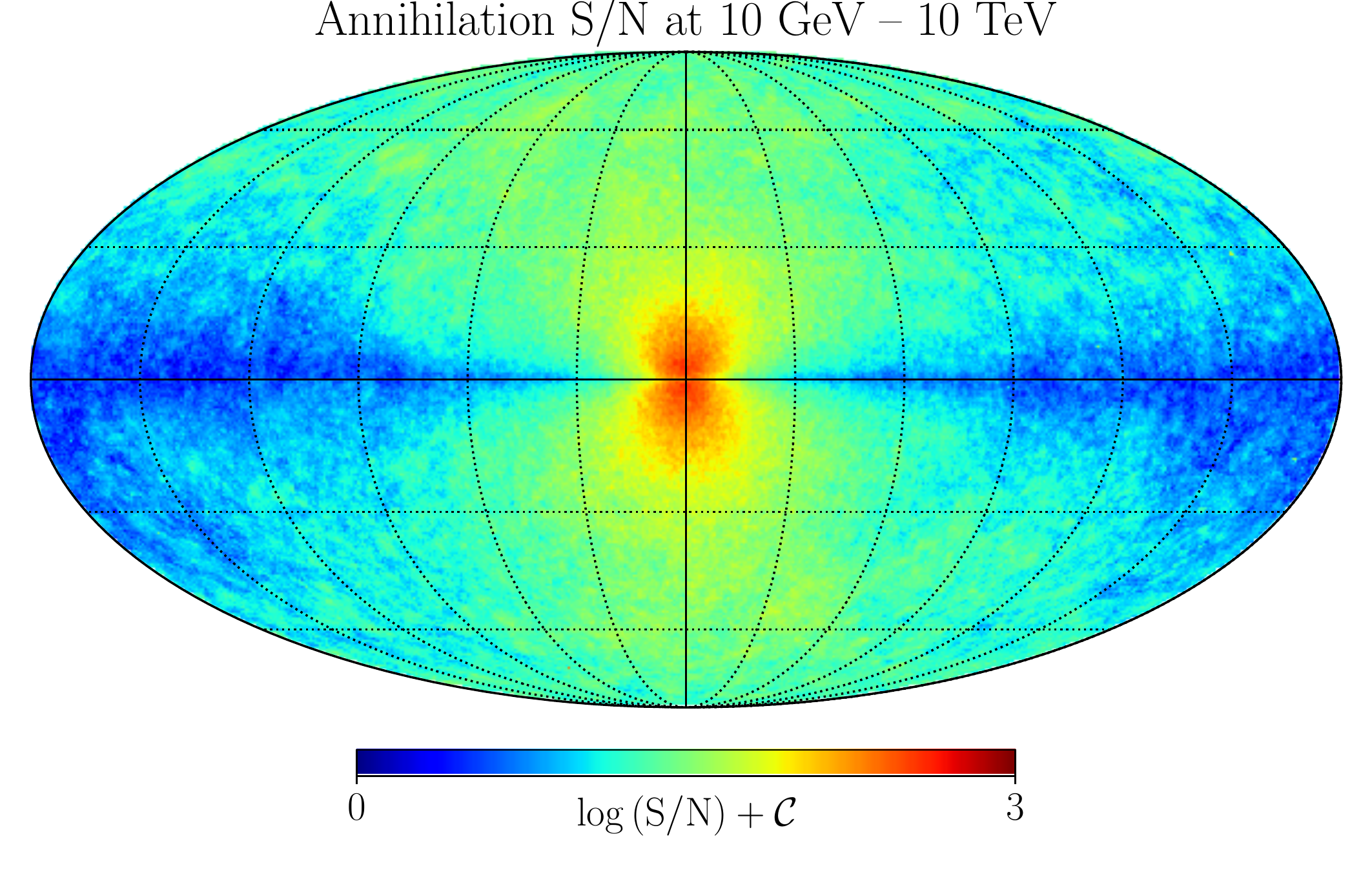}
    \hspace{10pt} \includegraphics[width=0.45\textwidth]{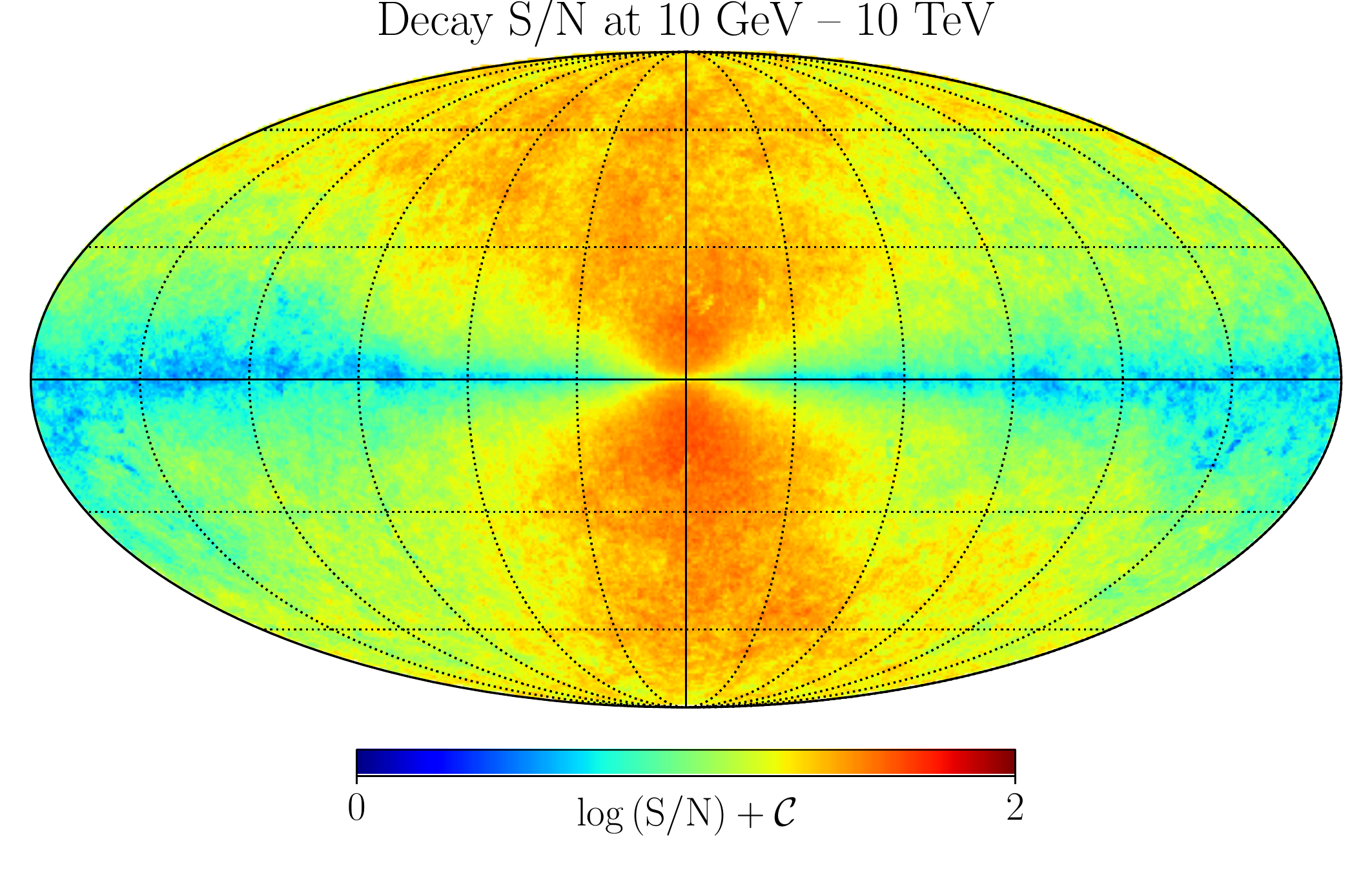}
    \end{center}
    \caption{The all-sky S/N prediction of DM gamma-ray signal at $10 \text{ MeV} - 10\text{ GeV}$ (top) and $10\text{ GeV} - 10\text{ TeV}$ (bottom) scale for annihilation (left) and decay (right) emission. Similar results can be found in Ref. \cite{Weniger:2012tx}, obtained with a DM analytic model and gamma-ray background provided by Fermi-LAT.}
    \label{fig:gsnmap}
\end{figure*}

\section{\label{app:level1}DM Flux and Spectroscopy}
In general, DM can produce photons via either annihilation or decay processes. The expected signal depends on the details of the DM particle model, the detector properties, the spatial distribution and also on the kinematics of DM in the Galaxy. For an observer at the position of the Sun, $\vec{R}_{\odot}$, the number $N_{\chi}$ of photons from DM annihilation/decay in the FOV $\Delta \Omega$ along the l.o.s., with exposure time $\Delta T$ and detector effective area $A_{\rm  eff }(E)$, should be proportional to:
\begin{equation} \label{eq:anni_photon}
    N_{\chi}^{\rm{dec.(ann.)}} \propto \frac{\Delta T}{4\pi}   \int_{\Delta \Omega} d\vec{s}\, \int dE \,  \rho_{\chi}^{(2)}\,\frac{dN_{\chi} }{dE}A_{\text{eff}}\, ,
\end{equation}
where $\rho_{\chi}$ denotes the DM density at distance $r(\vec{s}\,) = \sqrt{R^{2}_{\odot}+s^2+ 2 \, \vec{s} \cdot \vec{R}_{\odot}}$ from the GC, and $\vec{s}$ identifies the direction of the observer l.o.s. -- equivalently specified by the pair $(l,b)$ -- and $dN_{\chi}/dE$ is the  annihilation spectrum. Note that due to the relative motion of DM with respect to the observer, $dN_{\chi}/dE$ gets distorted according to:
\begin{equation} \label{eq:modify_spec}
   \frac{d N_{\chi}(E,l,b)}{dE} = \int d \mathcal{E} \frac{dN_{\chi}}{d \mathcal{E}} \mathcal{K}\left(\mathcal{E},E,\sigma_{los}[r(\vec{s}\,)] \right) \ ,
\end{equation}
where $\mathcal{K}$ is the kernel function that takes into account the Doppler-shift modification of the original DM spectrum and in simple cases may be approximated by a Gaussian profile, yielding for instance Eq.~\eqref{eq:DM_decay_Gauss} for DM decay; $\sigma_{los}$ is the DM velocity dispersion along the l.o.s.. More precisely, for an emission line centered at the rest-frame energy $E_{0}$, the energy redistribution from $\mathcal{K}$ is related to the Doppler effect $\Delta E = -v_{\text{los}}E_{0}/(c+v_{\text{los}})$ where $v_{\text{los}}$ is the l.o.s. velocity of DM relative to the observer.

Assuming an effective area constant in energy, and neglecting also the l.o.s. dependence in $\sigma_{los}$, both $A_{\rm eff}$ and the spectral integration reported in Eq.~\eqref{eq:anni_photon} can be factored out of Eq.~\eqref{eq:anni_photon}. Then, the observed spectrum from DM particles of mass $m_{\chi}$ and annihilation cross section $\langle \sigma v \rangle$ would actually be
\begin{eqnarray} \label{eq:anniflux}
    \frac{d\Phi_{\chi}(E,l,b)}{d\Omega dE} = 
    \frac{\langle \sigma v \rangle}{8\pi m_{\chi}^{2}} \frac{dN_{\chi}}{dE} \int d\vec{s} \, \rho_{\chi}^{2}  \, .
\end{eqnarray}
Following Ref.~\cite{diemand07,kuhlen08, springel08}, we can define the DM relative luminosity as
\begin{equation} \label{eq:anni_L}
    \mathcal{L}^{\text{ann.}}(l,b) = \int_{\text{los}} d\vec{s}  \, \rho_{\chi}^{2}[r(\vec{s}\,)] \, ,
\end{equation}
which only depends on the DM spatial distribution. Similarly, we expect from DM decay with rate $\Gamma_{\chi}$:
\begin{equation} \label{eq:decayflux}
    \frac{d\Phi_{\chi}(E,l,b)}{d\Omega dE} = \frac{\Gamma_{\chi}}{4\pi m_{\chi}} \frac{dN_{\chi}}{dE} \mathcal{L}^{\text{dec.}}  \, ,
\end{equation}
where the DM decay luminosity corresponds to
\begin{equation} \label{eq:decay_L}
    \mathcal{L}^{\text{dec.}}(l,b) = \int_{\text{los}} d\vec{s} \rho_{\chi}[r(\vec{s}\,)] \, .
\end{equation}

\begin{figure}
    \centering
    \includegraphics[width=0.45\textwidth]{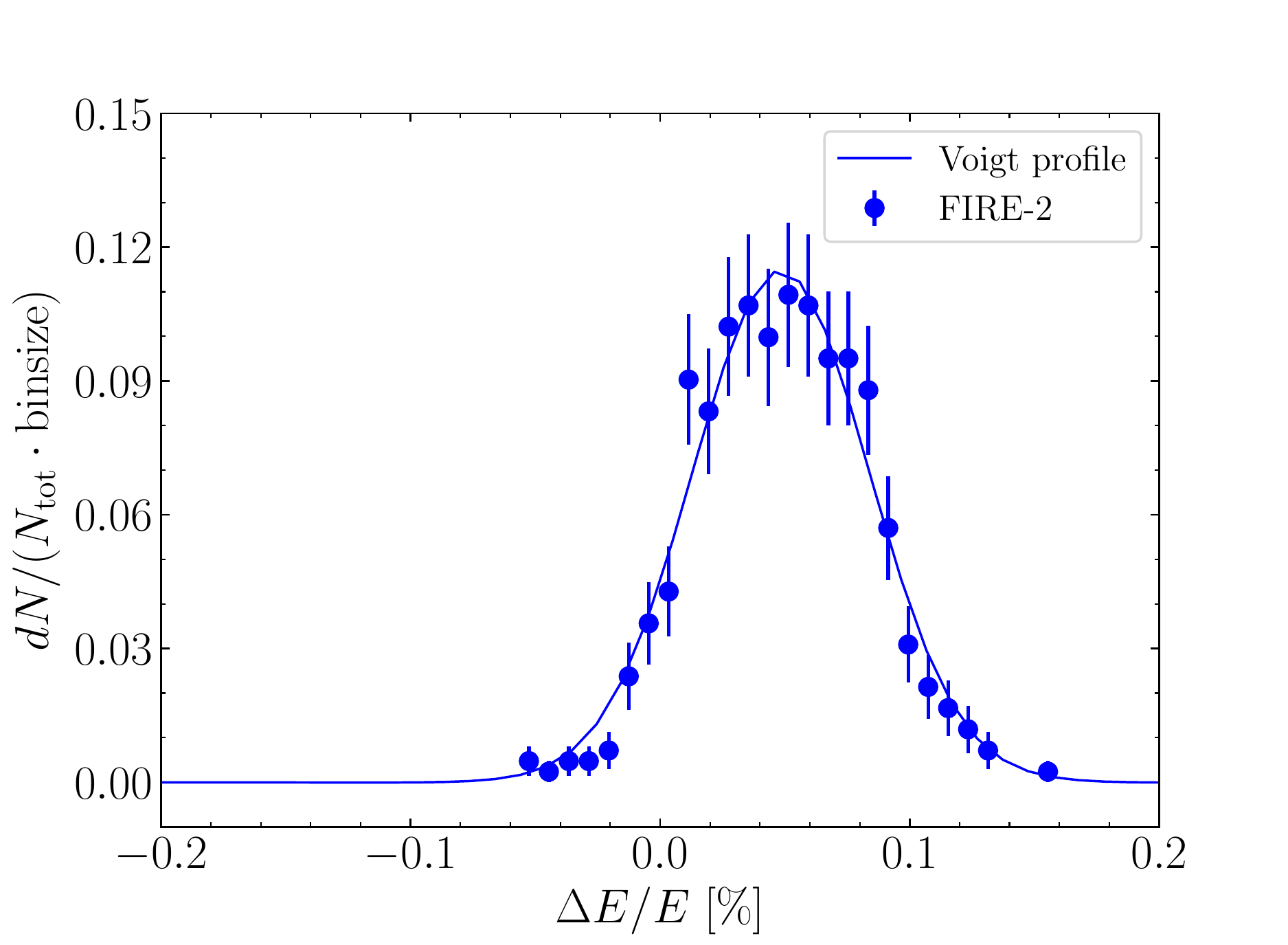}
    \includegraphics[width=0.45\textwidth]{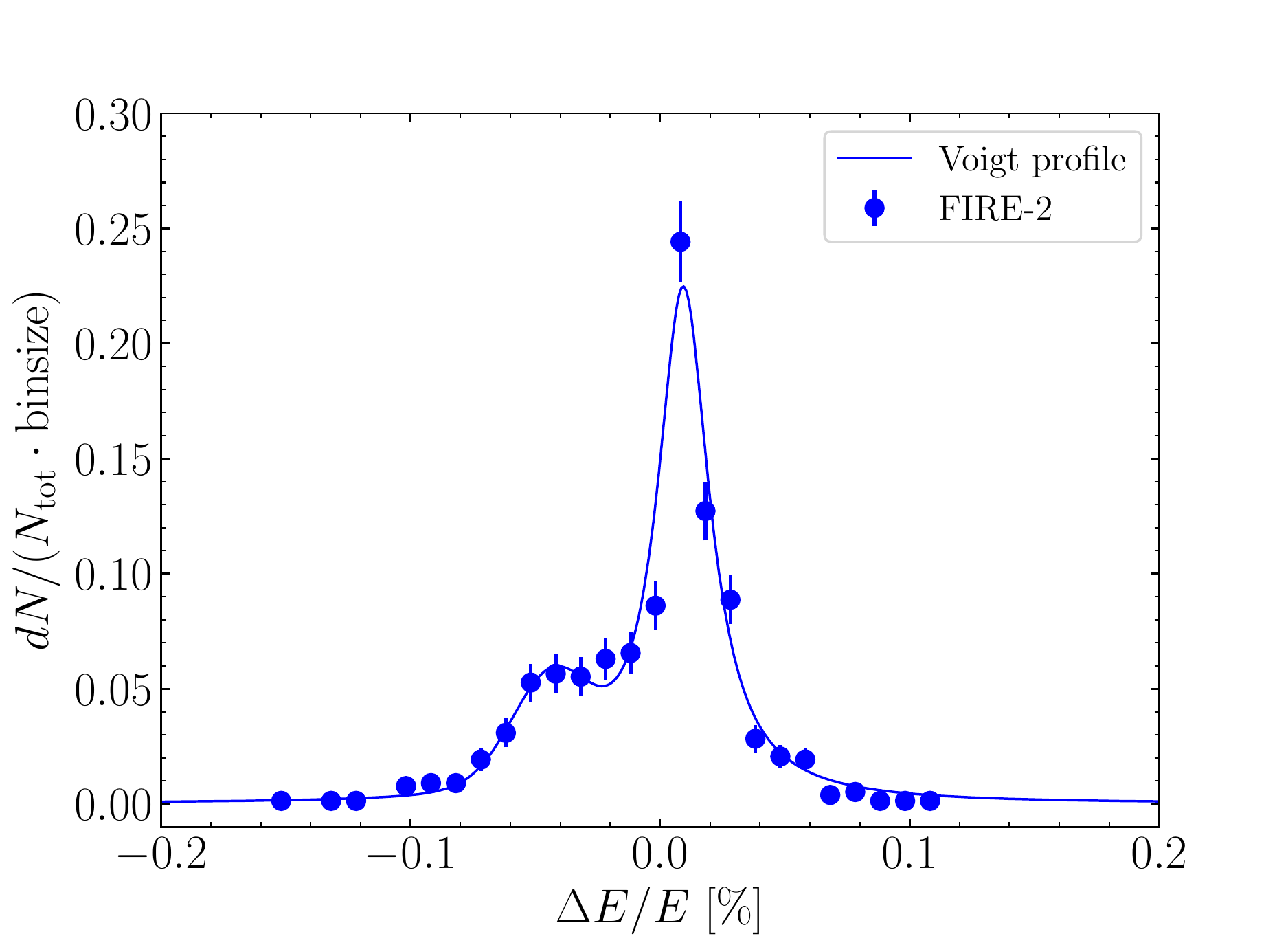}
    \caption{The Voigt profile fitting results for two line-of-sights with HEALPix ID 5136 (single Voigt function) and ID 7798 (two Voigt funtions). These two plots are for illustration purposes. Fitting with single Voigt function is sufficient for most cases (upper panel), while multiple line profiles are needed when subhalos contribute along the l.o.s. (lower panel). }
    \label{fig:voigt}
\end{figure}

\section{\label{app:level2}Fitting the Velocity Distribution Function}
In this appendix we briefly describe the procedure to extract the kernel function $\mathcal{K}$ from DM simulation data.
First of all, we analyzed the outcome of the simulation in the GC frame, where the observer position corresponds to $\vec{R}_{\odot} = (-8.2, 0, 0)$ kpc, with velocity $\vec{v}_{\odot} = (20.38, 224.718, 3.90)$ km/s. Through a proper coordinate transformation, we then moved to the rest frame of the observer, and then computed the DM l.o.s. velocity $v_{\textrm{los}}$ as the projection along the l.o.s. identified by $\vec{s}$, i.e. $v_{\text{los}} = \vec{v} \cdot \vec{s}$.
Note that a positive/negative $v_{\text{los}}$ here means an outward/inward movement of DM particles along the direction of $\vec{s}$. 

Hence, we adopted a HEALPix resolution of~6 in order to map DM particles along a given l.o.s. in the sky. We reconstructed the l.o.s. DM velocity distribution for each of the 49152 pixels at hand. This is the minimal sky resolution we found to be necessary in order to meaningfully analyze DM velocity distribution from the N-body simulation. Finally, having collected the info on $v_{\rm los}$, we built up histograms relative to the energy shift $\Delta E/E_{0}$:
\begin{equation}
    \Delta E/E_{0} = - v_{\rm los}/(c+v_{\rm los})\, ,
\end{equation}
where $E_0$ is the characteristic energy in the DM-emission rest frame; we grouped all particles along a given l.o.s. with $\Delta E/E_{0} = 0.8$ as a bin size and adopted the widely used Voigt function to fit the resulting histogram. Practically, we exploited a convenient normalization factor of $ 10^4$ for $\Delta E/E_{0}$ histograms in order to facilitate the fitting procedure, without incurring in any spurious effect. A couple of examples for the best-fitting result are shown in Fig.~\ref{fig:voigt}. For most cases, a single Voigt profile well described the histogram obtained, but two Voigt profiles were needed in some cases (see Fig.~\ref{fig:voigt}) due to the presence of substructure. Also, sometimes the bin size has been adjusted to further improve the fitting outcome when required.
The FWHM of the Voigt profile was obtained using the following expression \cite{Voigt_Width}:
\begin{equation}
    f_{V} = 1.069\tilde{\gamma} + \sqrt{0.8664\tilde{\gamma}^{2} + 8\tilde{\sigma}^{2}\ln2}\, ,
\end{equation}
where $\tilde{\sigma}$ and $\tilde{\gamma}$ are the Gaussian and Lorentz profile parameters estimated in the fit. For multiple profile cases, in a conservative fashion, we adopted the center of the narrow component as the energy shift and used the FWHM of the broad component as the profile width.

\bibliographystyle{apsrev4-1}
\bibliography{dm}
\end{document}